%% file: main.tex
\documentclass[preprint,12pt]{elsarticle}
\usepackage{geometry}
\usepackage{url}
\usepackage{graphicx,graphics}
\usepackage{fullpage}

\include{packages_and_notations}

\begin{document}
\begin{frontmatter}
\title{A Machine Learning Approach to Adaptive Covariance Localization}

  \author[labela]{Azam Moosavi\corref{cor1}}
  \author[labelb,labelc]{Ahmed Attia}
  \author[labela]{Adrian Sandu}
  \address[labela]{Computational Science Laboratory \\
  Department of Computer Science   \\
  Virginia Polytechnic Institute and State University \\
  2201 Knowledgeworks II, 2202 Kraft Drive, Blacksburg, VA 24060, USA \\
  Phone: 540-231-2193, Fax: 540-231-9218    \\
  E-mail: \href{azmosavi@vt.edu}{azmosavi@vt.edu}}
  \address[labelb]{Mathematics and Computer Science Division, Argonne National Laboratory, Argonne, IL}
  \address[labelc]{Statistical and Applied Mathematical Science Institute (SAMSI), RTP, NC}
  
\include{logo}
  %
  \begin{abstract}
    Data assimilation plays a key role in large-scale atmospheric weather forecasting, where the state of the physical system is estimated from model outputs and observations, and is then used as initial condition to produce accurate future forecasts.
    The Ensemble Kalman Filter (EnKF) provides a practical implementation of the statistical solution of the data assimilation problem and has gained wide popularity as. This success can be attributed to its simple formulation and ease of implementation.
    EnKF is a Monte-Carlo algorithm that solves the data assimilation problem by sampling the probability distributions involved in Bayes' theorem. Because of this, all flavors of EnKF are fundamentally prone to sampling errors when the ensemble size is small. In typical weather forecasting applications, the model state space has dimension $10^{9}-10^{12}$, while the ensemble size typically ranges between $30-100$ members. Sampling errors manifest themselves as long-range spurious correlations and have been shown to cause filter divergence. To alleviate this effect covariance localization dampens spurious correlations between state variables located  at a large distance in the physical space, via an empirical distance-dependent function. The quality of the resulting analysis and forecast is greatly influenced by the choice of the localization function parameters, e.g., the radius of influence. The localization radius is generally tuned empirically to yield desirable results. Optimal tuning of the localization function parameters is still an open problem.
	This work, proposes two adaptive algorithms for covariance localization in the EnKF framework, both based on a machine learning approach. The first algorithm adapts the localization radius in time, while the second algorithm tunes the localization radius in both time and space.
    Numerical experiments carried out with the Lorenz-96 model, and a quasi-geostrophic model, reveal the potential of the proposed machine learning approaches.
    %
  \end{abstract}
 \end{frontmatter}
\section{Introduction}
\label{Sec:Introduction}
%
Predicting the behavior of complex dynamical systems by computer simulation is  crucial in numerous fields such as oceanography, glaciology, seismology, nuclear fusion, medicine, and atmospheric sciences, including weather forecasting and meteorology. Data assimilation (DA) is the set of methodologies that combine multiple sources of information about a physical system, with the goal of producing an accurate description of the state of that system. 
These sources of information include computer model outputs, observable measurements, and probabilistic representations of errors or noise. DA generates a best representation of the state of the system, called the analysis, together with the associated uncertainty. In numerical weather prediction the analysis state that can be used to initialize subsequent computer model runs that produce weather forecasts.

DA algorithms generally fall into one of two main categories, variational and statistical approaches.
The variational approach for DA solves an optimization problem to generate an analysis state that minimizes the mismatch between model-based output and collected measurements, based on the level of uncertainty associated with each~\cite{attia2015hmcsmoother,attia2016reducedhmcsmoother,Kalnay_2002_book}.
Algorithms in the statistical approach apply Bayes' theorem to describe the system state using a probability distribution conditioned by all available sources of information.
A typical starting point for most of the algorithms in this approach is the Kalman filter (KF)~\cite{kalman1961new, kalman1960new}, which assumes that the underlying sources of errors are normally distributed, with known means and covariances.
Applying KF in large-scale settings is not feasible due to the intrinsic requirement of dealing with huge covariance matrices.

The EnKF ~\cite{Burgers_1998_EnKF, Evensen_1994, houtekamer1998data}
follows a Monte-Carlo approach to propagate covariance information, which makes it a practical approach for large-scale settings.
Various flavors of EnKF have been developed~\cite{Burgers_1998_EnKF,Evensen_1994,Evensen_2003,Houtekamer_1998a,ott2004local,Bishop_2001_ETKF,anderson2001ensemble}, and have been successfully used for sequential DA in oceanographic and atmospheric applications. 

EnKF carries out two main computational stages in every assimilation cycle,  while operating on an ensemble of the system states to represent probability distributions.
The ``forecast'' (prediction) stage involves a forward propagation of the system state from a previous time instance to generate an ensemble of forecast states at the current time. 
The ``analysis'' stage uses the covariance of the system states to assimilate observations and to generate an analysis ensemble, i.e., a Monte-Carlo representation of the posterior distribution of the model state conditioned by measurements.

In typical atmospheric applications the model state space has dimension $\sim 10^{9}-10^{12}$, and a huge ensemble is required to accurately approximate the corresponding covariance matrices. However, computational resources limit the number of ensemble members to $30-100$, leading to ``under-sampling''~\cite{houtekamer1998data} and its consequences: filter divergence, inbreeding, and long-range spurious correlations~\cite{anderson2013empirical,anderson2007exploring, anderson2012localization,petrie2008localization}.
A lot of effort has been dedicated to solving the issue of under-sampling. 
Inbreeding and the filter divergence are alleviated by some form of inflation~\cite{anderson1999monte}.
Long range spurious correlations are removed by covariance localization~\cite{hamill2001distance}.
Covariance localization is implemented by multiplying the regression coefficient in the Kalman gain
with a decaying distance-dependent function such as a Gaussian~\cite{anderson2012localization} or the Gaspari-Cohn fifth order piecewise polynomial ~\cite{Gaspari_1999_correlation}.

Different localization techniques have been recently considered for different observation types,
different type of state variables, or for an observation and a state variable that are separated in time. For horizontal localization, only a single GC function is used, and is tuned by finding the best value of its parameter 
(half the distance at which the GC function goes to zero). Tuning the localization for big atmospheric problems is very expensive.
Similar challenges appear in the vertical localization \cite{anderson2007exploring}.

Previous efforts for building adaptive algorithms for covariance localization includes the work of Anderson~\cite{anderson2007adaptive} based on a hierarchical ensemble filter. This approach adapts the localization radius by minimizing the root-mean-square difference of regression coefficients obtained by a group of ensemble filters. 
Adaptive localization by formulating the Kalman update step as a differential equation in terms of its ensemble members was proposed in \cite{bergemann2010mollified}, and computing localization as a power of smoothed ensemble sample correlation
was discussed in \cite{bishop2007flow, bishop2009ensemble}.
The approach proposed in~\cite{anderson2012localization} formulates localization as a function of ensemble size and correlation between the observation and state variable. Correlation factors are obtained and applied as traditional localization for each pair of observation and state variable during assimilation.
An Observing System Simulation Experiment (OSSE) algorithm is developed in~\cite{anderson2013empirical, lei2014comparisons}. 
OSSE computes the localization radius of influence from a set of observation-state pairs by minimizing the root-mean-square (RMS) of the posterior ensemble mean compared to true model state. 
%
A probabilistic approach proposed in~\cite{zhen2014probabilistic} defines the optimal radius of influence as the one that minimizes the distance between the Kalman gain using the localized sampling covariance and the Kalman gain using the true covariance. Further, the authors generalized this method for the case when the true covariance is unknown but it can be estimated probabilistically based on the ensemble sampling covariance.
The relation between the localization length for domain localization and observation localization is investigated in~\cite{kirchgessner2014choice}. This study concluded that the optimal localization length is linearly
dependent on an effective local observation dimension given by the sum of the observation weights.
In~\cite{lei2014comparisons} two techniques for estimating the localization function are compared. 
The first approach is the Global Group Filter (GGF) which minimizes the RMS difference between the estimated regression coefficients using a hierarchical ensemble filter. The second approach is the Empirical Localization Function (ELF) that minimizes the RMSE difference between the true values of the state variables and the posterior ensemble mean. 
The ELF has smaller errors than the hand-tuned filter, while the GGF has larger errors than the hand-tuned counterpart.

In this study we propose to adapt covariance localization parameters using machine learning algorithms. Two approaches are proposed and discussed. In the \textit{localization-in-time} method the radius of influence is held constant in space, but it changes adaptively from one assimilation cycle to the next. In the \textit{localization-in-space-and-time} method
the localization radius is space-dependent, and is also adapted for each assimilation time instance.
The learning process is conducted off-line based on historical records such as reanalysis data, and the trained model is subsequently used to predict the proper values of localization radii in future assimilation windows.

The paper is organized as follows. Section~\ref{Sec:Background} reviews the EnKF algorithm, the under-sampling issue and typical solutions, and relevant machine learning models. 
Section~\ref{Sec:Adaptive_Loc_Alg} presents the new adaptive localization algorithms in detail, and discusses their computational complexity and implementation details.
Section~\ref{Sec:Experimental} discusses the setup of numerical experiments and the two test problems, the Lorenz and the quasi-Geostrophic (QG) models. Numerical results with the adaptive localization methodology are reported in Section~\ref{Sec:Numerical_Results}. Conclusions and future directions are highlighted in Section~\ref{Sec:Conclusions}.

\section{Background}
\label{Sec:Background}
%
This section reviews the mathematical formulation of EnKF, and associated challenges such as under-sampling, filter divergence, and development of long-range spurious correlations. 
We discuss traditional covariance localization, a practical and successful ad-hoc solution to the problem of long-range spurious correlations, that requires an empirical tuning of the localization parameter, e.g., the radius of influence. 
%
The last subsection reviews the basic elements of a machine learning algorithm, with special attention being paid to random forests.
\subsection{Ensemble Kalman filters}
\label{Subsec:EnkF}
EnKF proceeds in a prediction-correction fashion and carries out two main steps in every assimilation cycle: \textit{forecast} and \textit{analysis}. Assume an analysis ensemble $\{\xa_{k-1}(e) \mid e = 1,\ldots,\Nens\}$ is available at a time instance $t_{k-1}$. In the forecast step, an ensemble of forecasts $\{\xf_k(e) \mid e = 1,\ldots,\Nens\}$ is generated by running the numerical model forward to the next time instance $t_k$ where observations are available:
\begin{subequations}
\label{eqn:EnKF_equations}
\begin{equation}
\label{eqn:EnKF_forecast}
\xf_k(e) = \mathcal{M}_{t_{k-1}\rightarrow t_k}(\xa_{k-1}(e)) + \eta_k(e),~ e=1,\ldots, \Nens \,,
\end{equation}
where $\mathcal{M}$ is a discretization of the model dynamics.
To simulate the fact that the model is an imperfect representation of reality, random model error realizations $\eta_k(e)$ are added. Typical assumption is that the model error is a random variable distributed according to a Gaussian distribution $\mathcal{N}(0,\mat{Q}_k)$. In this paper we follow a perfect-model approach for simplicity, i.e., we set $\mat{Q}_k = \mat{0}$ $\forall k$.

The generated forecast ensemble provides estimates of the ensemble mean $\xbarf_k $ and the flow-dependent  background error covariance matrix $\mat{B}_k$ at time instance $t_k$:
\begin{eqnarray}
\label{eq:EnKF_analysis}
\xbarf_k   &=&  \frac{1}{\Nens} \sum_{e=1}^{\Nens}{\xf_k(e) } \,, \\
\mat{X}^{'}_k   &=&  \bigl[\xf_k(e)- \xbarf_k\bigr]_{e=1,\ldots, \Nens} \,, \\
\mat{B}_k   &=& \frac{1}{\Nens-1}\,  \mat{X}^{'}_k \, \mat{X}^{'}_k\,^T , \quad
\mat{X}^{'}_k   =  \bigl[\xf_k(e)- \xbarf_k\bigr]_{e=1,\ldots, \Nens}.
\end{eqnarray}
In the analysis step each member of the forecast is analyzed separately using the Kalman filter formulas~\cite{Burgers_1998_EnKF,Evensen_1994}:
\begin{eqnarray}
\label{eqn:EnKF_Analysi_and_gain}
\xa_k(e)  &=&  \xf_k(e) + \mat{K}_k \left( \left[\yk + \zeta_k(e)\right] - \mathcal{H}_k(\xf_k(e)) \right),\ \\
 \label{eqn:EnKF_gain}
\mat{K}_k    &=&  \mat{B}_k \mat{H}^T_k { \left(\mat{H}_k \mat{B}_k \mat{H}^T_k + \mat{R}_k \right)}^{-1}\,,
\end{eqnarray}
\end{subequations}
where $\mat{H}_k = \mathcal{H}_k'(\overline{\x}^{\rm f}_k) $ is the linearized observation operator, e.g. the Jacobian, at time instance $t_k$. The stochastic (``perturbed'') version of the EnKF~\cite{Burgers_1998_EnKF} adds different realizations of the observation noise $\zeta_k \in \mathcal{N}(0,\mat{R}_k)$ to each individual observation in the assimilation procedure.  The same Kalman gain matrix $\mat{K}_k$ is used to assimilate observations(s) to each member of the forecast ensemble.

Deterministic (``square root'') versions of EnKF~\cite{Tippett_2003_EnSRF} avoid adding random noise to observations, and thus avoid additional sampling errors. They also avoid the explicit construction of the full covariance matrices and work by updating only a matrix of state deviations from the mean. A detailed discussion of EnKF and variants can be found in~\cite{Evensen_2007_book,asch2016data}. 
%

\subsection{Inbreeding, filter divergence, and spurious correlations}
EnKF is subject to sampling errors due to under-sampling whenever the number of ensembles is too small to be statistically representative of the large-dimensional model state. In practical settings, under-sampling leads to filter divergence, inbreeding, and the development of long-range spurious correlations~\cite{houtekamer1998data}. 

\paragraph{Inbreeding and filter divergence}
In inbreeding the background error is under-estimated, which causes the filter to put more emphasis on the background state and less emphasis on the observations. This means that the forecast state is influenced adequately by the observational data, 
and the filter fails to adjust an incorrectly estimated forecast estate. Inbreeding and the filter divergence can be resolved using covariance inflation \cite{anderson1999monte}; this is not further considered in this work. However, the machine learning approach proposed here for covariance localization can be extended to inflation.

\paragraph{Long-range spurious correlations}
The small number of ensemble members may result in a poor estimation of the true correlation between state components, or between state variables and observations. In particular, spurious correlations might develop between variables that are located at large physical distances, when the true correlation between these variables is negligible. As a result, state variables are artificially affected by observations that are physically remote ~\cite{anderson2001ensemble,hamill2001distance}. This generally results in degradation of the quality of the analysis, and eventually leads to filter divergence. 

\subsection{Covariance localization}
Covariance localization seeks to filter out the long range spurious correlations and enhance the estimate of forecast error covariance~\cite{hamill2001distance,houtekamer2001sequential,whitaker2002ensemble}.
Standard covariance localization is typically carried out by applying a Schur (Hadamard) product~\cite{million2007hadamard,schur1911bemerkungen} between a correlation matrix $\rho$ with distance-decreasing entries and the ensemble estimated covariance matrix, resulting in the localized Kalman gain:
\begin{equation} 
\label{eqn_loclized_gain}
\mat{K}_k = \left( \rho \circ \mat{B}_k \right) \mat{H}^T_k { \left(\mat{H}_k \left(\rho \circ  \mat{B}_k \right) \mat{H}^T_k + \mat{R}_k \right)}^{-1}\,,
\end{equation} 
Localization can be applied to $\mat{H}_k \mat{B}_k $, and optionally to the $\mat{B}_k$ projected into the observations space $\mat{H}_k \mat{B}_k \mat{H}_k^T$~ \cite{petrie2008localization}. Since the correlation matrix is a covariance matrix, the Schur product of the correlation function and the forecast background error covariance matrix is also a covariance matrix. 
Covariance localization has the virtue of increasing the rank of the flow-dependent background error covariance matrix $\rho \circ  \mat{B}_k$, and therefore increasing the effective sample size.

A popular choice of the correlation function $\rho$ is defined by the Gaspari-Cohn (GC) fifth order piecewise polynomial ~\cite{Gaspari_1999_correlation}  function that is non-zero only for a small local region and zero every other places \cite{petrie2008localization}: 
\begin{equation}
\label{eqn:gaspari_cohn}
\rho(z) =\left\{ 
\begin{array}{lr}
- \frac{1}{4} \left( \|z\| /c \right)^5 + \frac{1}{2} \left( \|z\|/c \right)^4 +  \frac{5}{8} \left( \|z\|/c \right)^3 - \frac{5}{3} \left( \|z\|/c \right)^2 + 1, \quad  & 0 \leq \|z\| \leq c, \\
\\
 \frac{1}{12} \left( \|z\| /c \right)^5 - \frac{1}{2} \left( \|z\|/c \right)^4 +  \frac{5}{8} \left( \|z\|/c \right)^3 + \frac{5}{3} \left( \|z\|/c \right)^2  \quad & c \leq \|z\| \leq 2c, \\
 - 5 \left( \|z\| /c \right)+4-\frac{2}{3} \left( c/\|z\| \right), \\
 \\
 0, \quad & 2c \leq \|z\| 
\end{array}
\right. 
\end{equation}
The correlation length scale is $c=\sqrt{\frac{10}{3}}\, \ell$, \cite{lorenc2003potential} where $\ell$ is a characteristic physical distance. The correlation decreases smoothly from $\rho(0)=1$ to zero at a distance more than twice the correlation length. Depending on the implementation, $z$ can be either the distance between an observation and grid point or the distance between grid points in the physical space.

\subsection{Machine learning and random forests} 
\label{Subsec:ML_and_random_forest}
Machine learning has found numerous applications in
 data science, data mining, and data analytics. However, the immense potential of applying machine learning to
help solve computational science problems remains largely untapped to date. Recently, data-driven approaches
to predict and model the approximation errors of low-fidelity
and surrogate models have shown promising results \cite{moosavi2017multivariate, wu2016physics}.  
The multivariate predictions of local reduced-order-model method (MP-LROM) \cite{moosavi2017multivariate} proposes
a multivariate model to compute the error of local reduced-order surrogates. 
In \cite{ attia2016clusterHMC} a new filter in DA frame work is developed which is called Cluster Hybrid Monte Carlo sampling filter (CLHMC) for non-Gaussian data assimilation which relaxes the Gaussian assumption in the original HMC filter by employing a clustering step. 

One of the fundamental algorithms in ML is \emph{regression analysis}. Generally speaking, multivariate regression models~\cite{freedman2009statistical} approximate the relationship  between a set of dependent variables, and a set of independent variables. multivariate dependent variables.  We review next a popular ML approach for multivariate regression that incorporates \emph{random forests} for model fitting, and then apply it to adaptive covariance localization.

\subsubsection{Random forests}
 \label{Susubbsec:random_forest}

Random forests (RFs) are ensemble-based learning methods~\cite{breiman2001random, breiman1996bagging} based on the idea  that \textit{a group of weak learners can come together to form a strong learner}. The final decision or model is then built by some sort of averaging over the group of week learners forming a strong learner~\cite{breiman1996bagging, dietterich2000ensemble}. RFs can efficiently solve various ML problems including classification and regression. RFs are superior to other similar algorithms and can run efficiently on large datasets~\cite{breiman2001random}. Amongst the most special-purpose popular versions of RFs are Iterative Dichotomiser 3 (ID3)~\cite{quinlan1986induction} and it's successor (C4.5)~\cite{quinlan2014c4}, and conditional inference trees~\cite{strobl2008conditional}.

Specifically, RFs work by constructing an ensemble of decision trees, such that each tree builds a classification or regression model in the form of a tree structure. Instead of using the whole set of features available for the learning algorithm at once, each subtree uses a subset of features. The ensemble of trees is constructed using a variant of the bagging~\cite{breiman1996bagging, dietterich2000ensemble} (bootstrap aggregation) technique, thus yielding a small variance of the learning algorithm~\cite{dietterich2000ensemble}. Furthermore, to ensure robustness of the ensemble-based learner, each subtree is assigned a subset of features selected randomly in a way that minimizes the correlation between individual learners.

While RFs can efficiently handle large data sets with high feature dimensionality, they can suffer from over-fitting, a general problem faced by almost all machine learning algorithms. To avoid over-fitting in an RF one has to optimize the tuning parameter that governs the feature split, and the number of features assigned to each tree from the bootstrapped data~\cite{segal2004machine}.
Consider a dataset $D$ and a set of features $F$ to be used by the RF. For each tree in the forest, a bootstrap sample $D^i\subset D$ is randomly selected.  Instead of examining all possible feature-splits, a subset of the features $f \subset F$ with $|f| \ll |F|$ is randomly selected~\cite{liaw2002classification}. Each node then splits on the best feature in the subset $f$ rather than $F$. This approach has the advantage that the RFs can be efficiently constructed in parallel, and that the correlation between trees in the ensemble is reduced. Random sampling and bootstrapping can be efficiently applied to RFs to generate a parallel, robust, and very fast learner for high-dimensional data and features. 
%
%

\section{Machine Learning Approach for Adaptive Localization}
\label{Sec:Adaptive_Loc_Alg}
%
This section develops two machine learning approaches for adaptive covariance localization. In the first approach the localization radius changes in time, meaning that the same localization radius is used at all spatial points, but at each assimilation cycle the localization radius differs. In the second approach the localization radius changes both in time and space, and is different for each assimilation cycle and for each state variable. In both approaches the localization radius is adjusted depending on the model behavior and overall state of the system. Here we study what features of the solution affect the most the optimal value of localization radius, such that using that localization radius the difference between analysis and the true state gets minimized.
The random forest approach is used to construct the learning model that takes the impactful set of features as input and outputs the localization radius. We now describe in detail the features and the objective function of the proposed learning model.


\subsection{Features and decision criteria}
\label{Subsec:features_and_decision_crit}

ML algorithms learn a function that maps input variables $F$, the features of the underlying problem, onto output target variables. The input to the learning model is a set of features $F$ which describes the underlying problem. During the learning phase the algorithm finds the proper function using a known data set. This function is used to predict target outputs given new instances of input variables. In this work the target variables are the localization radii at each assimilation cycle. We consider atmospheric models that have numerous variables and parameters, and select the feature set $F$ that capture the characteristics of the important behavioral patterns of the dynamical system at each assimilation cycle.  Specifically, the idea is to focus on the set features that best reflect the changes in analysis state with respect to changes in the localization radii.
 
%
\paragraph{Selection of the feature set}
We now consider the selection of important features of the model results and data sets to be fed into the ML algorithms. 
Relying on the Gaussianity assumption of the prior distribution, natural features to consider are the first and second order moments of the prior distribution of the model state at each assimilation cycle. However, the large dimensionality of practical models can make it prohibitive to include the entire ensemble average vector (forecast state $\xf$) as a feature for ML.  
One idea to reduce the size of the model state information is to select only model states with negligible correlations among them, e.g., states that are physically located at distances larger than the radius of influence. Another useful strategy to reduce model features is to select descriptive summaries such as the minimum and the maximum magnitude of state components in the ensemble.

The correlations between different variables of the model are descriptive of the behavior of the system at each assimilation cycle, and therefore are desirable features. Of course, it is impractical to include the entire state error covariance matrix among the features. Following the same reasoning as for state variables, we suggest including blocks of the correlation matrix for variables located nearby in physical space, i.e., for  subsets of variables that are highly correlated.
%

\paragraph{Decision criteria}

Under the Gaussianity assumption the quality of the DA solution is given by the quality of its first two statistical moments. Each of these aspects is discussed below.

A first important metric for the quality of ensemble-based DA algorithms is how well does the analysis ensemble mean (analysis state) represent the true state of the system. To quantify the accuracy of the ensemble mean we use the root mean-squared error (RMSE), defined as follows:
\begin{equation}
\label{eqn:RMSE}
	RMSE_k = \frac{1}{\sqrt{\Nstate}} \norm{\x_k - \xtrue(t_k)}_2 \,,
\end{equation}
where $\x^{\rm true}$ is the true system state, and $\norm{\cdot}_2$ is the Euclidian norm. 
Since the true state is not known in practical applications we also consider the deviation of the state from collected measurements as a useful indication of filter performance. 
The observation-state $RMSE$ is defined as follows:
\begin{equation}
\label{eqn:RMSE-obs}
	RMSE^{\x | \obs}_k = \frac{1}{\sqrt{\Nobs}} \norm{\mathcal{H}\left( \x_k \right) - \obs_k}_2 \,.
\end{equation}
Replacing $\x$ in ~\eqref{eqn:RMSE} and \eqref{eqn:RMSE-obs} with the forecast state $\xf$ or with the analysis state $\xa$ provides the formulas for the forecast or the analysis error magnitudes, respectively. The quality of the DA results measured by either \eqref{eqn:RMSE} in case of perfect problem settings, or by \eqref{eqn:RMSE-obs} in case of real applications. 

{\it In this work we use the observation-analysis error metric~\eqref{eqn:RMSE-obs}, denoted by $RMSE^{\xa | \obs}$, as the first decision criterion.}

A second important aspect that defines a good analysis ensemble is its spread around the true state. The spread can be visually inspected via the Talagrand diagram (rank histogram)~\cite{anderson1996method, candille2005evaluation}. A quality analysis ensemble leads to a rank histogram that is close to a uniform distribution. Conversely, U-shaped and Bell-shaped rank histograms  correspond to under-dispersion and over-dispersion of the ensemble, respectively. Ensemble based methods, especially with small ensemble sizes, are generally expected to yield U-shaped rank histograms, unless they are well-designed and well-tuned.
The calculation of the rank statistics in model space requires the ordering the true state entries with respect to the generated ensemble members, which is not feasible in practical applications. A practical rank histogram can alternatively be constructed by ordering the entries of the observation vector with respect to the ensemble members entries projected into the observation space~\cite{anderson1996method, candille2005evaluation}.

{\it In this work we use the uniformity of the analysis rank histogram, in observation space, as the second decision criterion.}

We now discuss practical ways to quantify the level of uniformity of rank histograms. The level of uniformity of forecast rank histograms is used as a learning feature, and that of the analysis histogram as a decision criterion.

A reasonable approach is to quantify the level of uniformity by the similarity between a distribution fitted to the rank histogram and a uniform distribution. A practical measure of similarity between two probability distributions $P$ and $Q$ is the Kullback-Leibler (KL) divergence~\cite{kullback1951information}:
\begin{equation}
\label{eqn:KL_genral}
\DKL{ P\| Q } = \mathbb{E}_{P} \left[ \log{(P)} - \log{(Q)} \right] \,.
\end{equation}
We first fit a beta distribution $Beta(\alpha, \beta)$ to the rank histogram (where the histogram domain $[0, \Nens]$ is mapped to $[0, 1]$ by a linear transformation). Considering that  $Beta(1, 1)$ is a uniform distribution over the interval $[0,1]$, we use the following measure of uniformity of the rank histogram:
\begin{equation}
\label{eqn:spread-metric}
 \DKL{ Beta(\alpha, \beta) \| Beta(1, 1) } \,.
\end{equation}
%

\subsection{ML-based adaptive localization algorithm}

We have identified two useful, but complementary, decision criteria, one that measures the quality of ensemble mean, and the second one that measures the quality of the ensemble spread. For the practical implementation we combine them into a single criterion, as follows:
\begin{equation}
\label{eqn:decision_criterion}
\mathcal{C}_{ \vec{r}} = w_1\, RMSE^{\xa | \obs} \, + w_2\, \DKL{ Beta(\alpha, \beta) \| Beta(1.0, 1.0) } \,,
\end{equation}
where the weighting parameters realize an appropriate scaling of the two metrics. The weights $w_1, w_2$ can be predefined, or can be learned from the data them as part of the ML procedure.

{\it The best set of localization radii are those that that minimize the combined objective function~\eqref{eqn:decision_criterion}.}
 
Algorithm~\ref{Alg:adaptive_localization} summarizes the proposed adaptive localization methodology.
\begin{algorithm}[h]
  \begin{algorithmic}[1]
   \State $dataset=[]$
    \State For $k \in \text{assimilation times}$
    \State $\quad$If $k \in \text{training  phase}$
    \State $\quad \quad Cost =\infty$
    \State $\quad \quad Radii\_Pool:= \bigl[ \vec{r_1}, \vec{r_2}, \vec{r_3}, \cdots, \vec{r_n} \bigr]$ 
    \State $\quad \quad$For $ i = 1$ to $n$
    \State $\quad \quad  \quad r=	Radii\_Pool [i]$
    \State $\quad \quad \quad$ Obtain $\x_k^f, \x_k^a $ 
    \State $\quad \quad \quad$Evaluate  $C_{\vec{r}}$ \eqref{eqn:decision_criterion}
    \State $\quad \quad \quad$If  $\left( C_{\vec{r}} < \textnormal{Cost} \right)$ Then
    \State  $\quad \quad \quad \quad Winner\_Radius= r$
    \State  $\quad \quad \quad $End If
    \State $\quad \quad  $End For loop
    \State $\quad \quad $Obtain $\x_k^f, \x_k^a$
    \State $\quad \quad dataset[K,:]=[Features, Winner\_Radius]$
    \State  $\quad$End If 
    \State $\quad  $Train the learning model with $dataset$
    \State $\quad $If $k \in \text{test  phase}$
    \State  $\quad \quad  r =$ Learning model predicts the localization radius 
    \State $\quad \quad$Perform localization with $r $
    \State $\quad \quad $Obtain $\x_k^f, \x_k^a$
    \State $\quad $End If 
    \State End For loop
  \end{algorithmic}
  \caption{Adaptive localization algorithm}
  \label{Alg:adaptive_localization}
\end{algorithm}
%
\subsection{Adaptive localization in time}
\label{Subsec:adaptive_time}
The first proposed learning algorithm uses the same (scalar) localization radius for all variables of the model. The value of this radius changes adaptively from one assimilation cycle to the next.  Specifically, at the current cycle we perform the assimilation using all localization radii from the pool and for each case compute the cost function $C_r$ \eqref{eqn:decision_criterion}.  After trying all possible radii from the pool, the radius associated with the minimum cost function is selected as winner. The analysis of the current assimilation cycle is the one computed using the winner radius. 

At each assimilation cycle we collect a sample consisting of the features described in \ref{Subsec:features_and_decision_crit} as inputs, and the winner localization radius $r$ as output (target variable) of the learning model. During the training phase, at each assimilation cycle, the ML algorithm learns the best localization radius corresponding to the system state and behavior. During the test phase, the learning model uses the current system information to estimate the proper value of the localization radius, without trying all possible values of localization radii.  Algorithm \ref{Alg:adaptive_localization} summarizes the adaptive localization procedure.  
%
\subsection{Adaptive localization in time and space}
\label{Subsec:adaptive_time_space}
The second proposed learning algorithm is to adapt the localization radii in both time and space. A different localization radius is used for each of the state variables, and these radii change at each assimilation cycle.
Here the localization radius is a vector $\vec{r}$ containing a scalar localization parameter for each state variable of the system. The pool of radii in this methodology contains multiple possible vectors. The pool can be large since it can include permutations of all possible individual scalar radii values. Similar to previous learning algorithm, at each time point we perform the assimilation with one of the vectors from the pool of radii, and select the one corresponding to the minimum cost function
as the winner.

At each assimilation cycle we collect a sample consisting of the model features as inputs and the winner vector of localization radii  as output of the learning model. In the training phase, the model learns the relation between system state and localization radii and during the test phase it estimates the proper value of localization radius for each state individually. The number of target variables the learning model could be as large as the number of state variables. This situation can be improved by imposing that the same scalar radii are used for multiple components, e.g., for entire areas of the physical model.

%

\subsection{Computational considerations}
During the training phase, the learning phase of the proposed algorithm needs to try all possible radii from the pool, and re-do the assimilation with that localization radius. This is computationally demanding, but the model can be trained off-line using historical data. The testing phase the learning model predicts a good value of the localization radius, which is then used in the assimilation; no additional costs are incurred except for the (relatively inexpensive) prediction made by the trained model.


\section{Setup of the Numerical Experiments}
\label{Sec:Experimental}
%
In order to study the performance of the proposed adaptive localization algorithm we employ two test models, namely the Lorenz-96 model~\cite{lorenz1996predictability}, and the QG-1.5 model~\cite{sakov2008deterministic}. 
All experiments are implemented in Python using the~\dates framework~\cite{attia2017dates}. The performance of the proposed methodology is compared against the deterministic implementation of EnKF (DEnKF) with parameters empirically tuned as reported in~\cite{sakov2008deterministic}. 

\subsection{Lorenz-96 model}
%
The Lorenz-96 model is given by \cite{lorenz1996predictability}:
\begin{equation}
  \label{eq:Base_Lorenz}
  \frac{dX_k}{dt}=-X_{k-2}X_{k-1}+X_{k-1}X_{k+1}-X_k+F,  \quad k=1,2, \cdots, K,
\end{equation}
with $K=40$ variables, periodic boundary conditions, and a forcing term $F=8$. 
A vector of equidistant component values ranging from $[-2,2]$ was integrated forward in time for $1000$ steps, each of size $0.005$ [units], and the final state was taken as the reference initial condition for the experiments. The background uncertainty is set to $8 \%$ of average magnitude of the reference solution. 
All state vector components are observed, i.e., $\mathcal{H} = \mat{I} \in \mathbb{R}^{K \times K}$ with $\mat{I}$ the identity operator. 
To avoid filter collapse, the analysis ensemble is inflated at the end of each assimilation cycle, with the inflation factor set to $\delta=1.09$.

\subsection{Quasi-geostrophic (QG-1.5) model}
\label{subsec:QG_Model}
The QG-1.5 model described by Sakov and Oke~\cite{sakov2008deterministic} is a numerical approximation of the following equations:
\begin{equation}
  \label{eqn:QG_Model}
  \begin{aligned}
    q_t &= \psi_x - \varepsilon J(\psi, q) - A \Delta^3 \psi + 2 \pi \sin(2 \pi y) \,, \\
    q & = \Delta \psi - F \psi \,,  \\
    J(\psi, q) & \equiv \psi_x q_x - \psi_y q_y \,, \\
    \Delta &:= \partial^2/\partial x^2 + \partial^2/\partial y^2 \,,
  \end{aligned}
\end{equation}
where $\psi$ is surface elevation or the stream function and $q$ is the potential vorticity.
We use the following values of the model coefficients ~\eqref{eqn:QG_Model} from \cite{sakov2008deterministic},:
$F=1600$, $\varepsilon=10^{-5}$, and $A=2\times 10^{-12}$. Boundary conditions used are $\psi = \Delta \psi = \Delta^2 \psi = 0$.
The domain of the model is a $1 \times 1$ [space units] square, with $0\leq x \leq 1,\, 0\leq y \leq 1$,  and is discretized by a grid of size $129 \times 129$ (including boundaries).
A standard linear operator to observe $300$ components of $\psi$ is used~\cite{attia2016clusterHMC}. The observation
error variance is $4.0$ [units squared] and synthetic observations are obtained by adding white noise to measurements of the see height level (SSH) extracted from a model run with lower viscosity~\cite{attia2016clusterHMC}. Here, the inflation factor is set to $\delta=1.06$, and the localization function is GC \eqref{eqn:gaspari_cohn} with the empirically-tuned optimal radius $\ell=3$.

\section{Numerical Results}
\label{Sec:Numerical_Results}

\subsection{Lorenz model with adaptive localization in time}
\label{Subsec:lor_adaptive_time}

The adaptive localization in time approach uses the same localization radius for all variables, and adapts the localization radius value at each assimilation cycle. This experiment has 100 assimilation cycles, where the first $80\%$ are dedicated to the training phase and the last $20\%$ to the testing phase.  The pool of radii for this experiment covers all possible values for the Lorenz model: $r \in [1, 40]$. 
 
\begin{figure}[h]
\centering
\subfigure[Training phase]{
\includegraphics[width=0.47\linewidth]{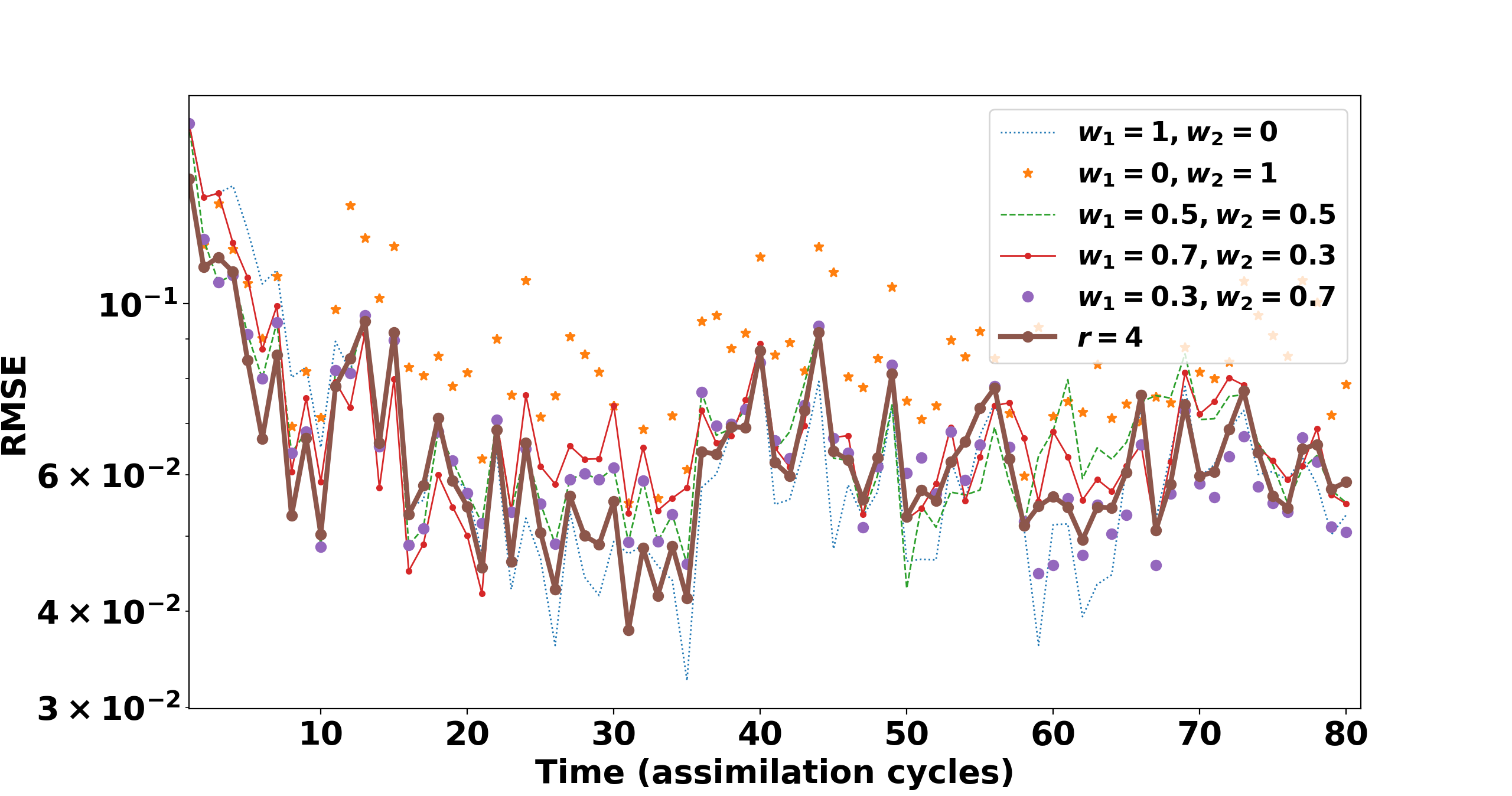}
\label{fig:lorenz_short_fixed_train}
}
\hfill
\subfigure[Testing phase]{
\includegraphics[width=0.47\linewidth]{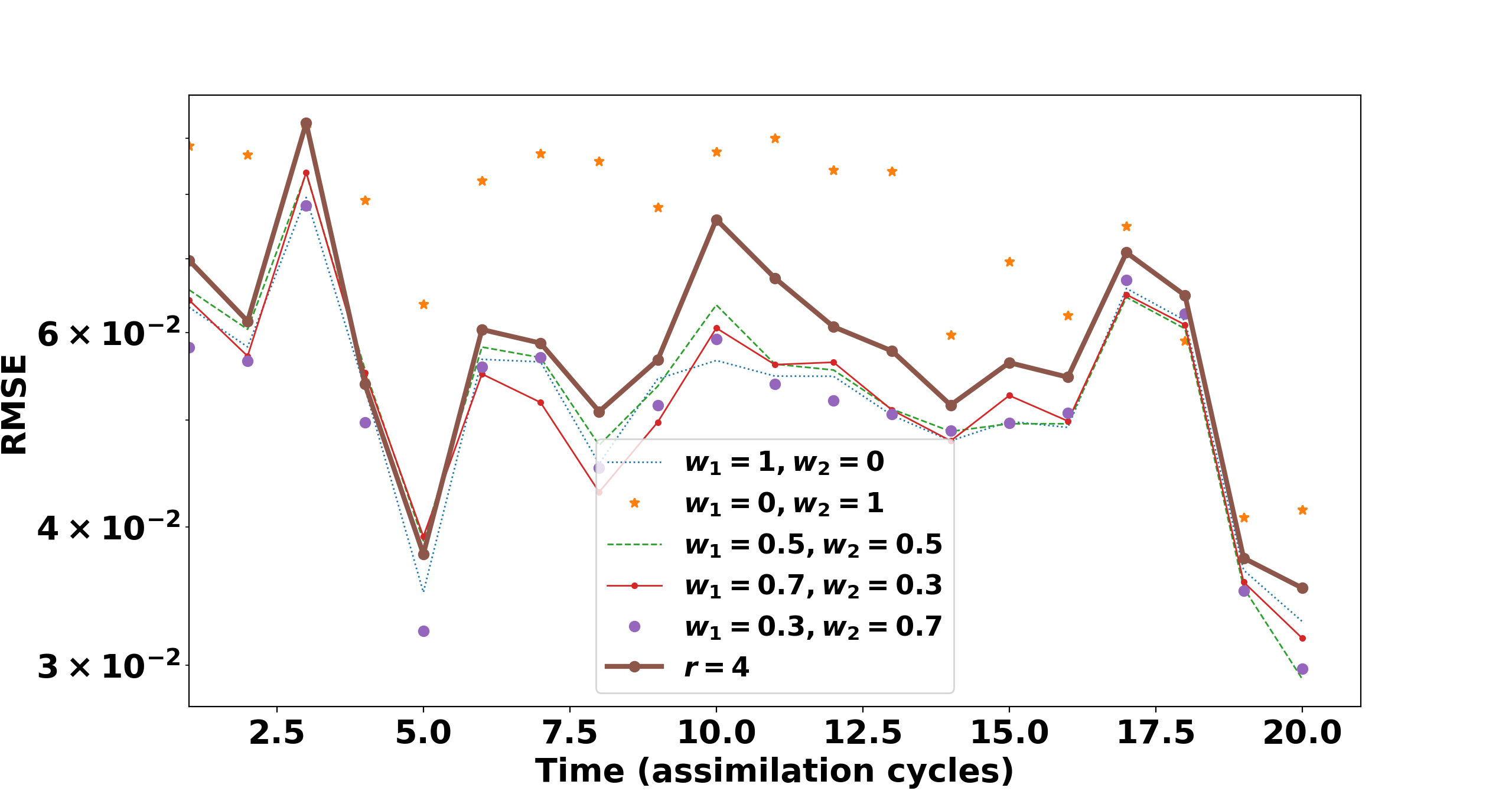}
\label{fig:lorenz_short_fixed_test}
}
\caption{Data assimilation with the Lorenz-96 model \eqref{eq:Base_Lorenz}. EnKF is applied with a Gaussian covariance localization function. EnKF results with adaptive covariance localization are shown for different choices of the weighting factors $w_1$, $w_2$. The localization is adaptive in time, and is compared to results with fixed localization radius. The training phase consists of 80 assimilation cycles, followed by the testing phase with 20 assimilation cycles. {\it The overall performance of the adaptive localization is better than that of the hand-tuned radius.}}
\label{fig:lorenz_short_fixed}
\end{figure}

We compare the performance of the adaptive localization algorithms against the best hand-tuned fixed localization radius value of $4$ which is obtained through testing all possible localization radii $([1, 40])$. Figure \ref{fig:lorenz_short_fixed} shows the logarithm of RMSE between analysis ensemble and the 
true (reference) state. 
The performance of adaptive localization methodology is evaluated for different weights $w_1$, $w_2$. The results indicate that increasing the weight of the KL distance measure increases the performance. {\it For the best choices of the weights the overall performance of the adaptive localization is slightly better than that of the fixed, hand-tuned radius.}

\begin{figure}[h]
  \begin{centering}
    \includegraphics[width=0.49\linewidth]{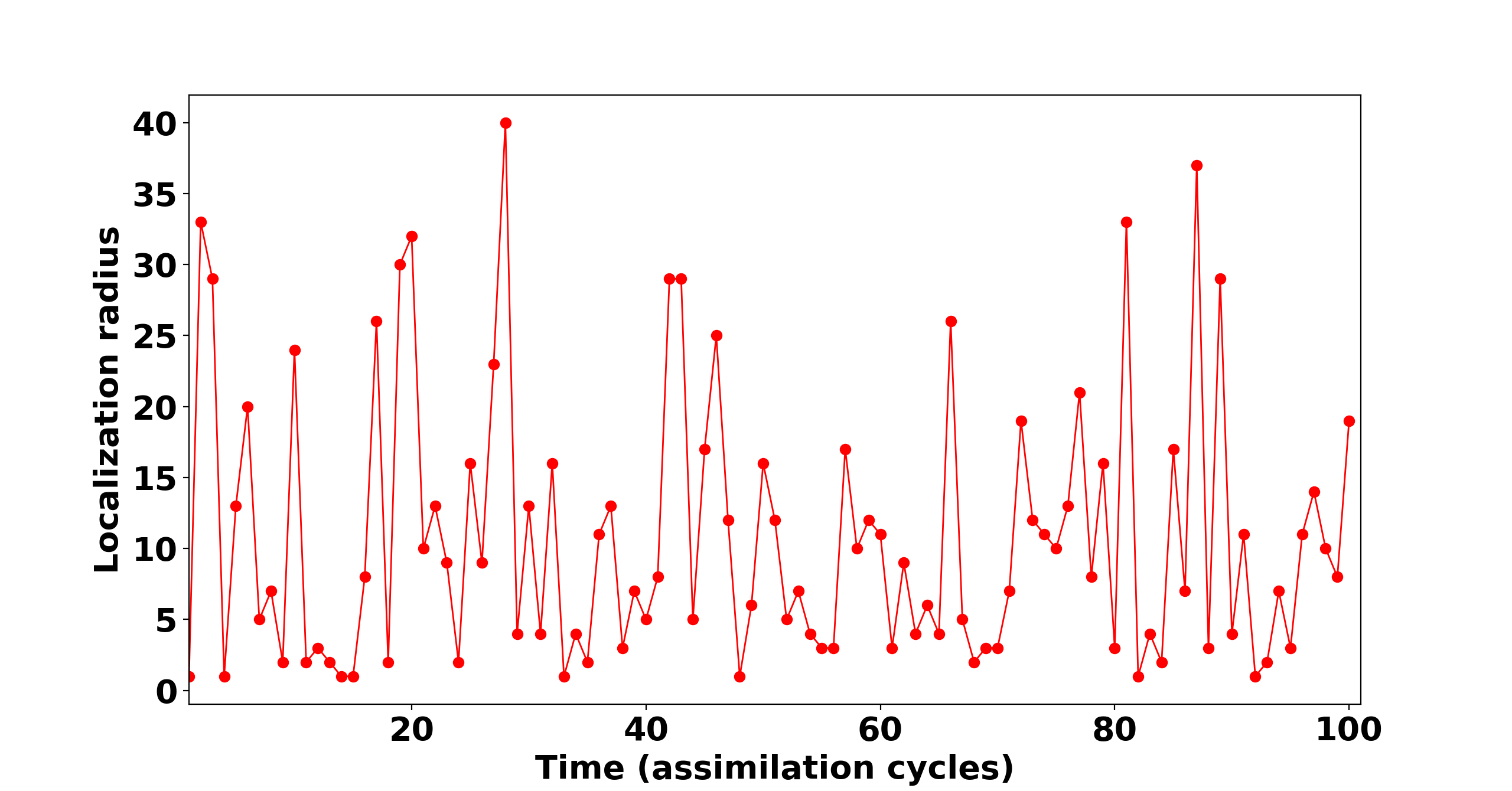}
    \caption{Data assimilation with the Lorenz-96 model \eqref{eq:Base_Lorenz}. Shown is the evolution of the localization radius in time over all 100 assimilation cycles. The weights of the adaptive localization criterion are $w_1=0.7$ and $w_2=0.3$.}
    \label{fig:radii_change_lorenz}
  \end{centering}
\end{figure}

Figure \ref{fig:radii_change_lorenz} shows the variability in the tuned localization radius over time for  both training and test phase. The weights of the adaptive localization criterion are $w_1=0.7$ and $w_2=0.3$ for this experiment. The adaptive algorithm changes the radius considerably over the simulation.
%

\begin{figure}[h]
  \begin{centering}
    \includegraphics[width=1.1\linewidth]{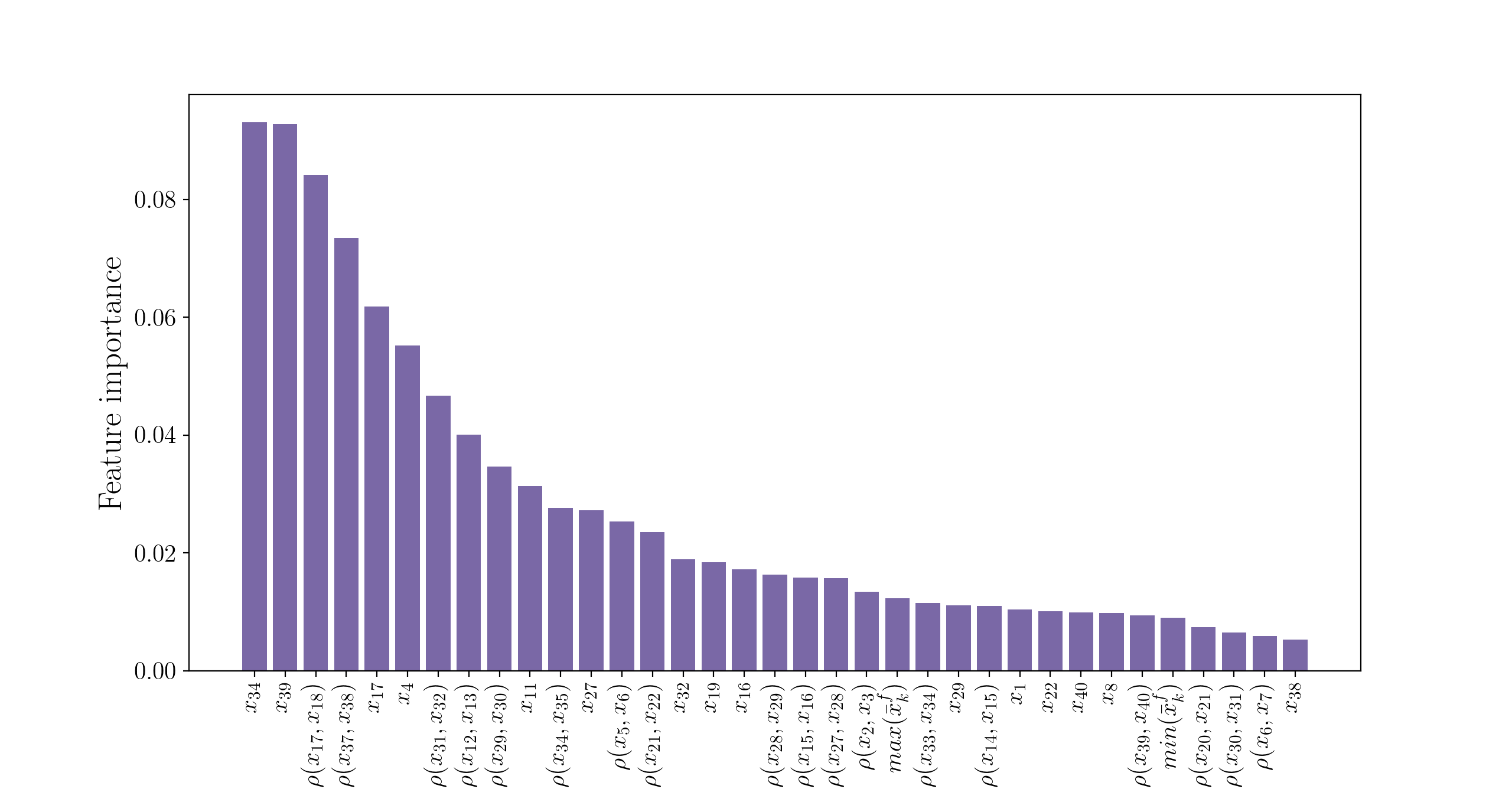}
    \caption{The 35 most important features of the Lorenz-96 model \eqref{eq:Base_Lorenz}}
    \label{fig:feature_importance_lorenz}
  \end{centering}
\end{figure}

Using the RF methodology we were able to recognize and select the most important features affecting the target variable prediction. Figure \ref{fig:feature_importance_lorenz} shows the 35 most important features of the Lorenz model which we included in our experiments.
%

\subsection{QG model  with adaptive localization in time}
\label{Subsec:qg_adaptive_time}
We use 100 assimilation cycles of the QG model, with $80\%$ dedicated to the training phase, and $20\%$ to the test phase. The pool of radii for this experiment is $r \in [1, 10]$.  EnKF is used with  $25$ ensemble members, inflation factor $\delta=1.06$, and GC localization function. An empirically optimal localization radius with these configurations was found by hand-tuning to be $r=3$. We use it as a comparison benchmark for the performance of the adaptive localization. 

\begin{figure}[h]
\centering
\subfigure[Training phase]{
\includegraphics[width=0.47\linewidth]{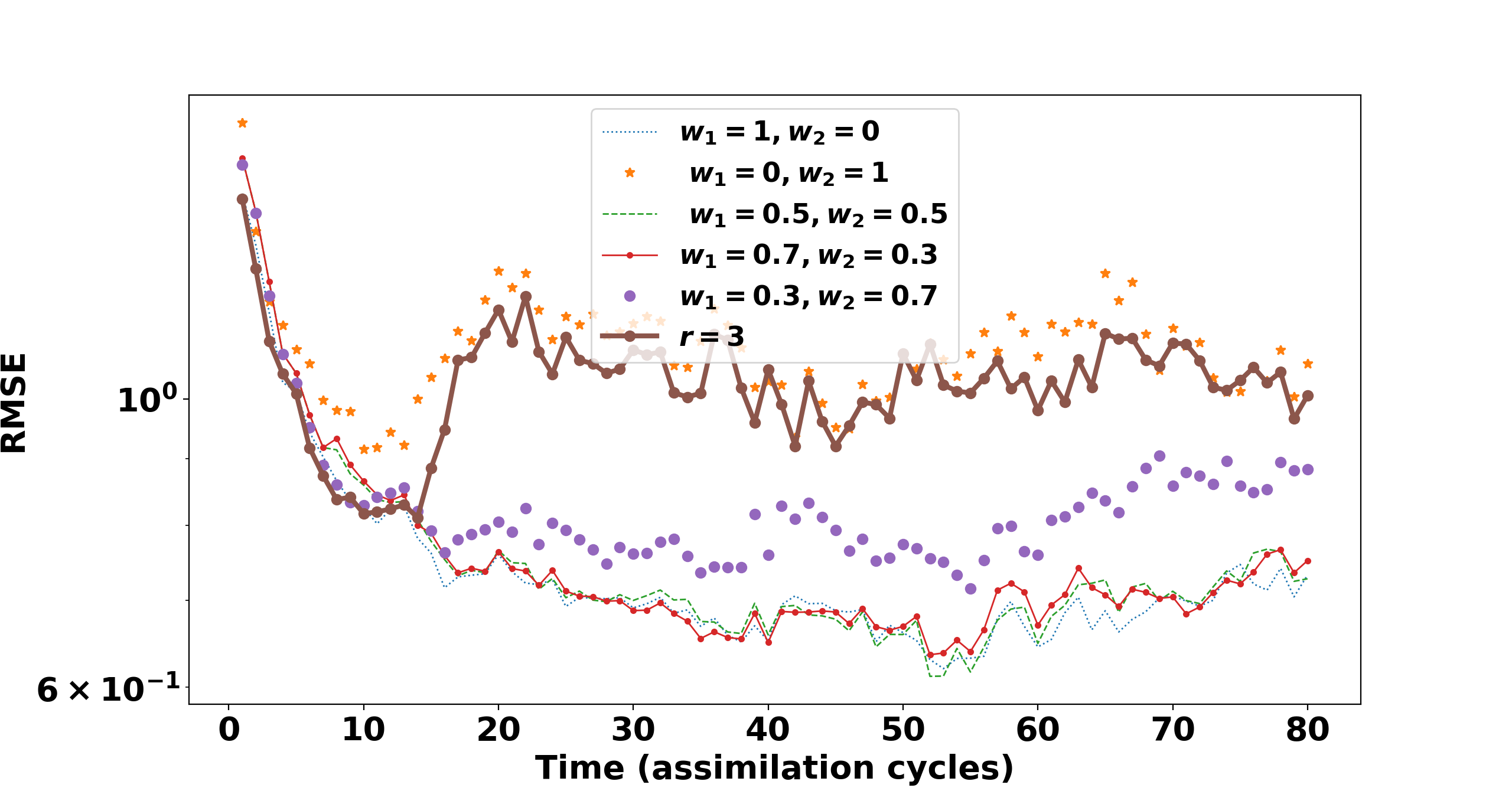}
\label{fig:QG_short_fixed_train}
}
\subfigure[Testing phase]{
\includegraphics[width=0.47\linewidth]{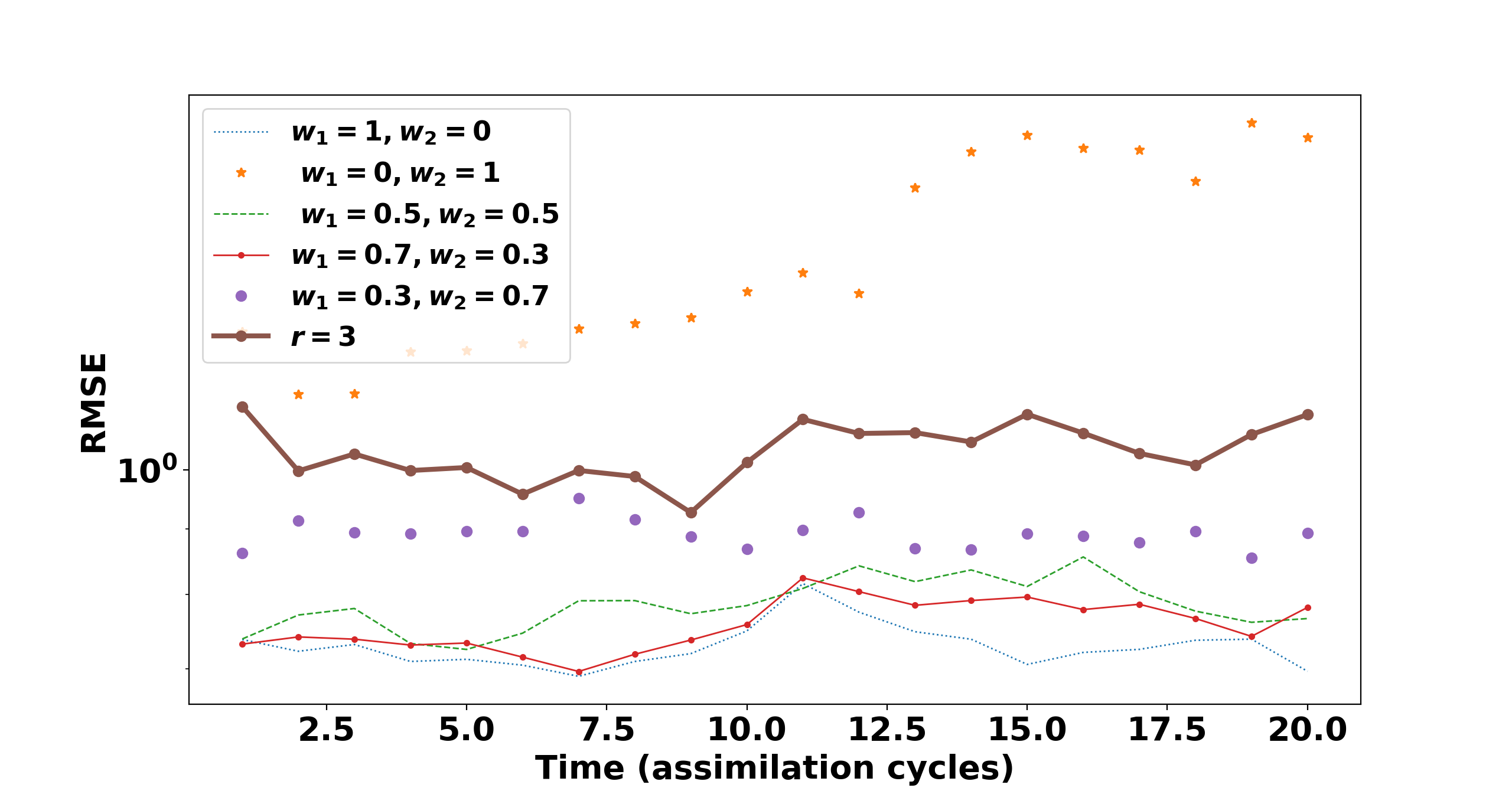}
\label{fig:QG_short_fixed_test}
}
\caption{Data assimilation with the QG model \eqref{eqn:QG_Model}. EnKF is applied with a GC covariance localization function. The (\textit{log}) RMSE is shown on the vertical axis, and the time (assimilation cycles) is shown on the horizontal axis. EnKF results with adaptive covariance localization are shown for different choices of the weighting factors $w_1,\, w_2$. The localization is adaptive in time, and is compared to results with fixed localization radius. The training phase consists of 80 assimilation cycles, followed by the testing phase with 20 assimilation cycles. {\it The analysis results with adaptive localization outperform those obtained with the hand-tuned radius.} }
\label{fig:QG_short_fixed}
\end{figure}

\begin{figure}[h]
  \begin{centering}
    \includegraphics[width=0.49\linewidth]{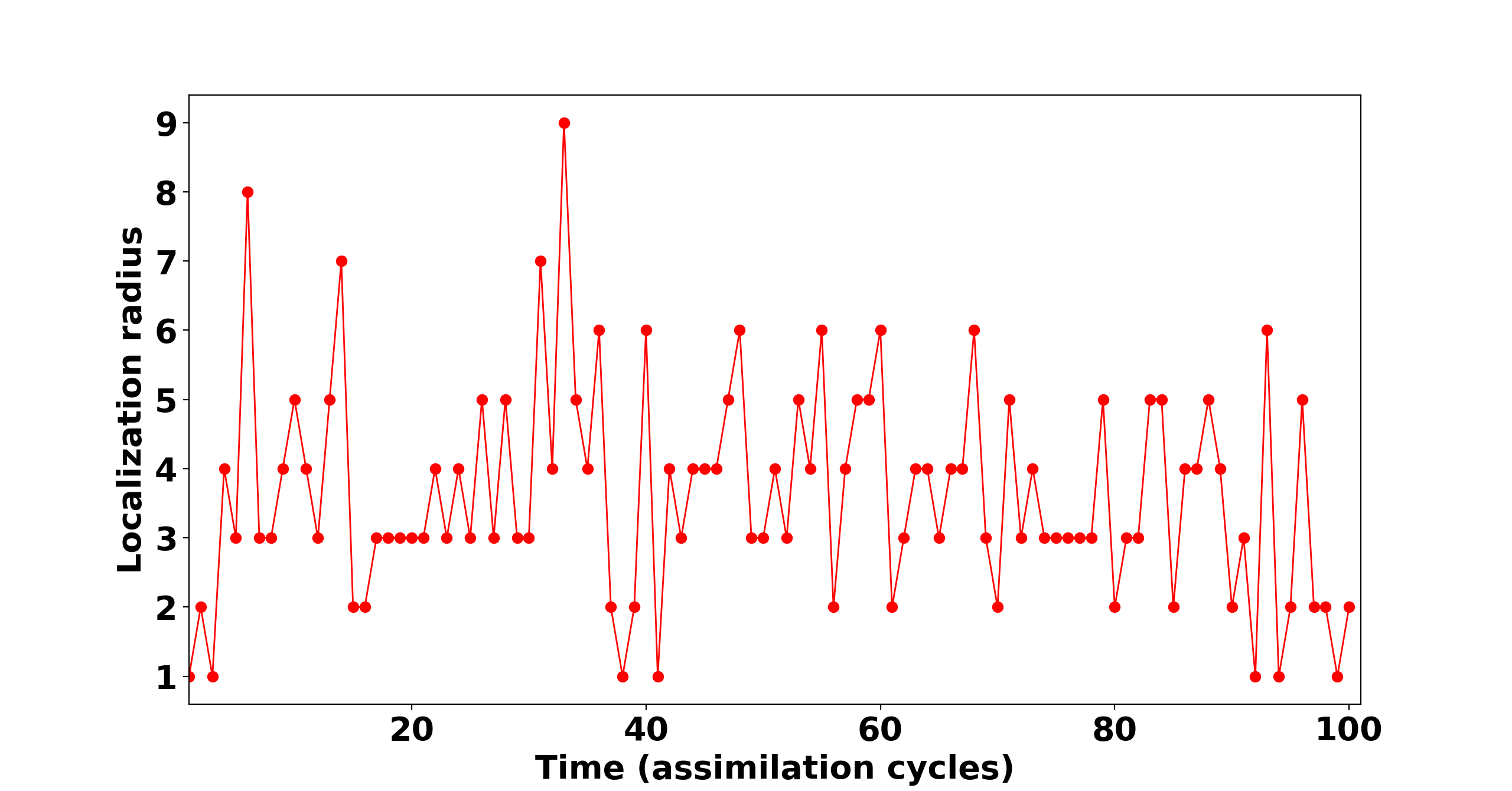}
    \caption{Data assimilation with the QG model \eqref{eqn:QG_Model}. Shown is the evolution of the localization radius in time over all 100 assimilation cycles. The weights of the adaptive localization criterion are $w_1=0.7$ and $w_2=0.3$}
    \label{fig:radii_change_QG}
  \end{centering}
\end{figure}

Figure \ref{fig:QG_short_fixed} shows the logarithm of RMSE between analysis obtained at each assimilation cycle and the true analysis. The performance of adaptive localization with  different weights $w_1$, $w_2$ is evaluated against the fixed localization with radius $r=3$. With higher weights for the KL distance measure, the performance of adaptive localization also increases. {\it The analysis results with adaptive localization outperform those obtained with the hand-tuned radius.} 
Figure \ref{fig:radii_change_QG} shows the variability of the localization radius in time for the weights $w_1=0.7$ and $w_2=0.3$. The adaptive algorithm changes the radius considerably over the course of the simulation.

\begin{figure}[h]
  \begin{centering}
    \includegraphics[width=1.1\linewidth]{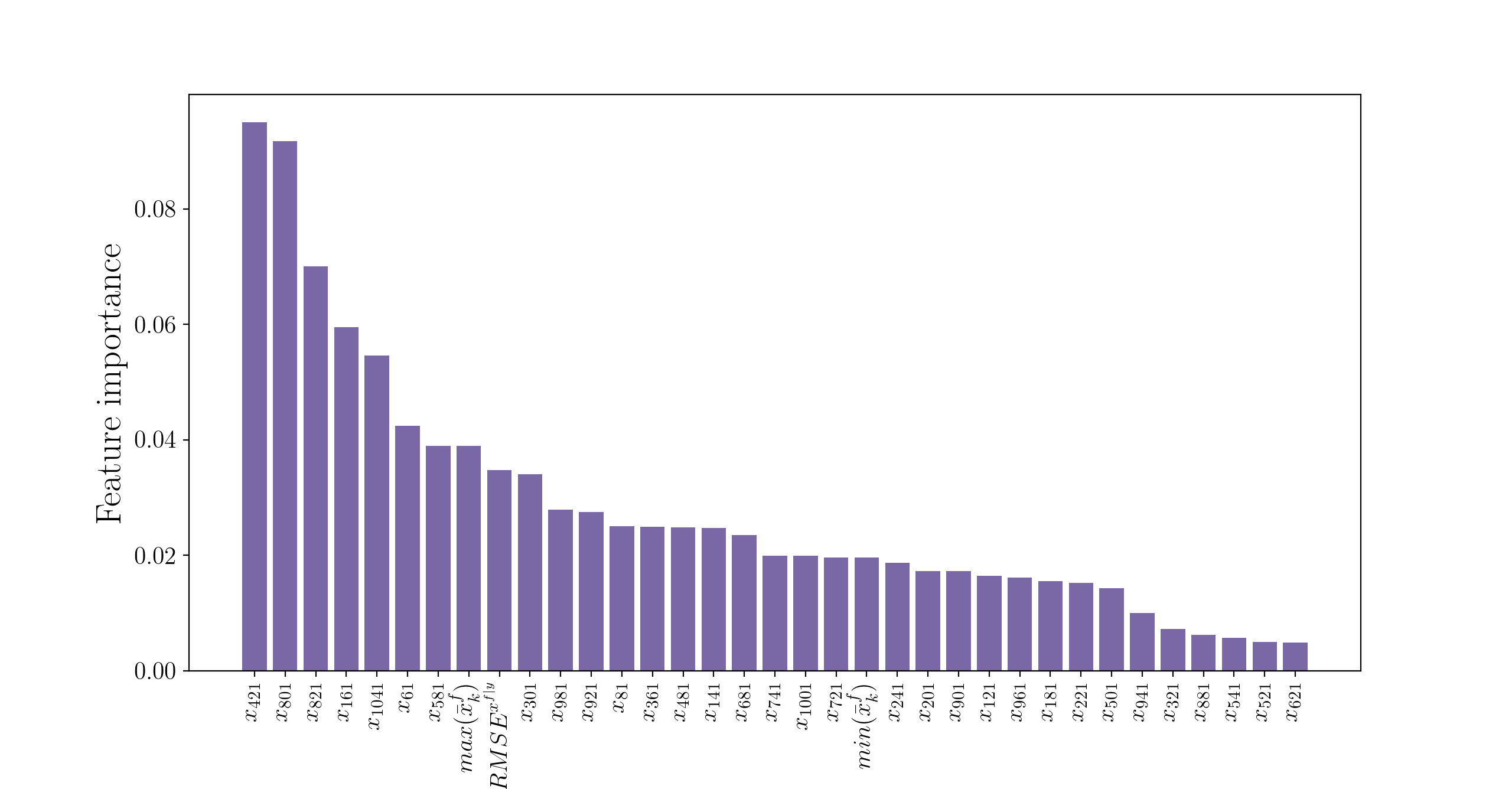}
    \caption{The 35 most important features of the QG model \eqref{eqn:QG_Model}}
    \label{fig:feature_importance_QG}
  \end{centering}
\end{figure}

Figure \ref{fig:feature_importance_QG} shows the 35 most important features of the QG model which we included in our experiments.
%

\subsection{Lorenz model  with adaptive localization in time and space}
\label{Subsec:lor_adaptive_time_space}

Here the localization radius changes at each assimilation cycle and it is also changing for each individual state variable of the system. The pool of radii for this experiment consists of random vectors of size $40$ where each component of the vector can have value in the range of all possible radii for the Lorenz model i.e $[1,40]$. Each component of the vectors in the pool can have different permutations of values in the range of $[1,40]$. The total number of  all possible vectors is huge, and testing all  in the training phase is infeasible. One way to limit the number of trials is to test randomly selected vectors of radii in the pool. For this experiments we set the number of trials to $30$ and at each trial we randomly pick a vector of radii from the pool. The localization radius of each state variable is the corresponding component in the  vector, the cost function of using each of the trials is obtained at each assimilation cycle. The number of target variables to estimate at each assimilation cycle in the test phase
is $40$ and hence we need more samples for the training phase.  The number of assimilation cycles for this experiment is $1000$, from which $80\%$ dedicated to the training phase, and $20\%$  to the testing phase. The EnKF uses $25$ ensemble members, the inflation factor of $1.09$ and the localization function is Gaussian.

\begin{figure}[h]
\centering
\subfigure[Training phase]{
\includegraphics[width=0.47\linewidth]{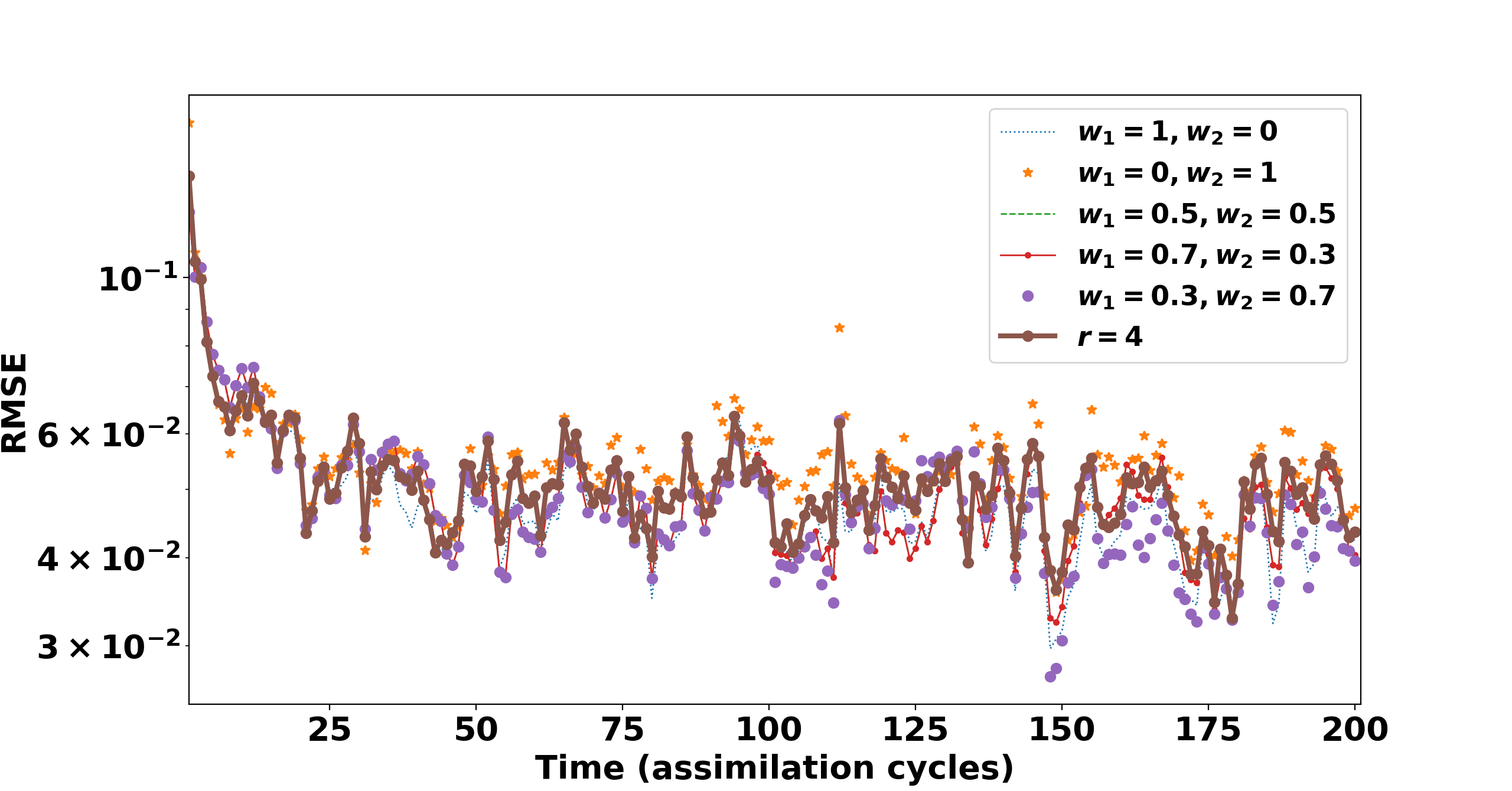}
\label{fig:lorenz_long_changing_train}
}
\subfigure[Testing phase]{
\includegraphics[width=0.47\linewidth]{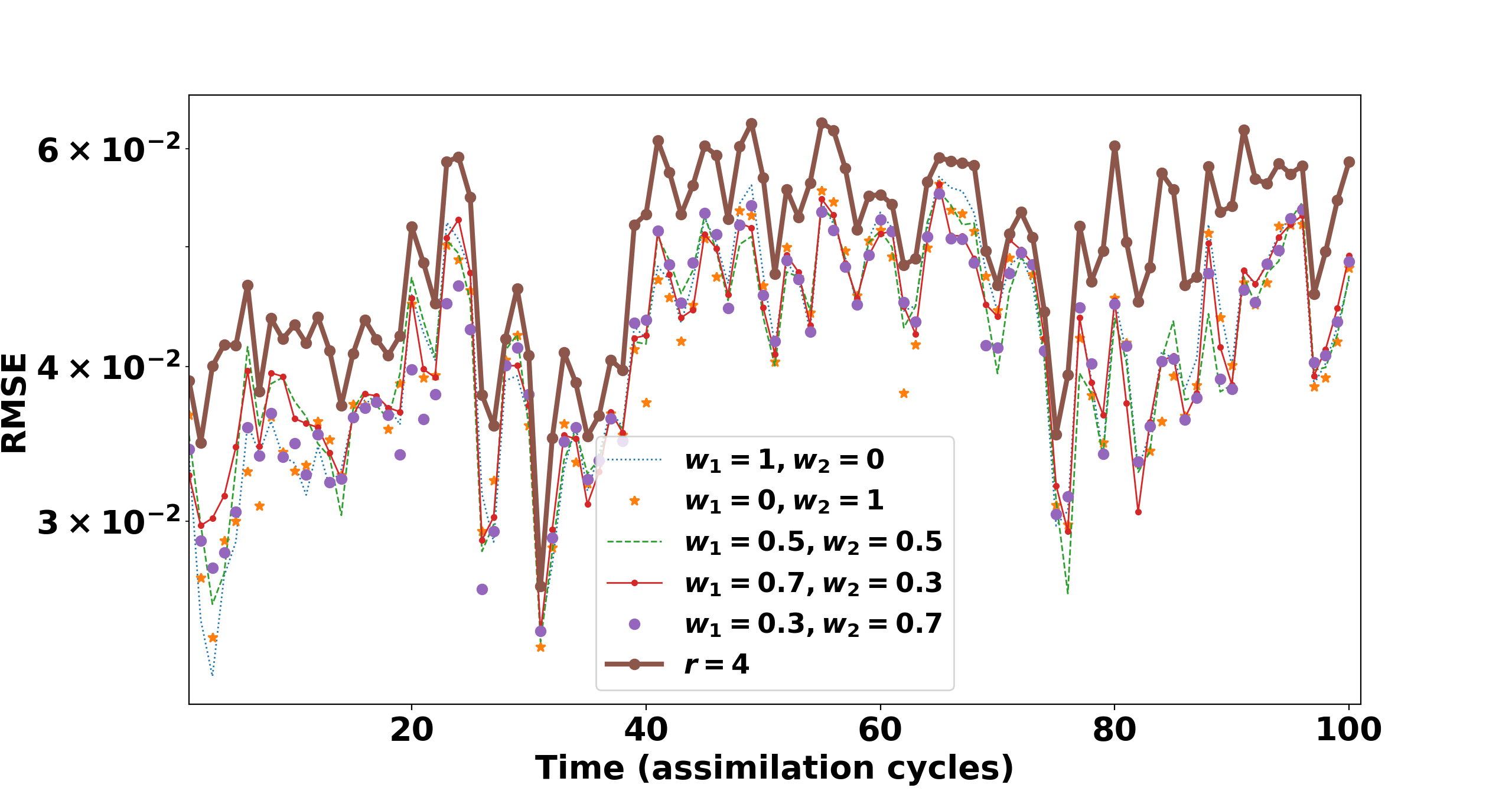}
\label{fig:lorenz_long_changing_test}
}
\caption{Data assimilation with the Lorenz-96 model \eqref{eq:Base_Lorenz}. EnKF is applied with a Gaussian covariance localization function. EnKF results with adaptive covariance localization are shown for different choices of the weighting factors $w_1$, $w_2$. The localization is adaptive in time and space, and is compared to results with fixed localization radius. The training phase consists of 800 assimilation cycles, followed by the testing phase with 200 assimilation cycles.}
\label{fig:lorenz_long_changing}
\end{figure}

Figure \ref{fig:lorenz_long_changing} shows the logarithm of RMSE between analysis and the reference state.  The performance of adaptive localization with different weights $w_1$, $w_2$ is evaluated against the fixed localization radius $r=4$. In the testing phase the results with the adaptive radii are slightly better than those with the optimal fixed radius.

\begin{figure}[h]
\centering
\subfigure[Average of localization radii]{
\includegraphics[width=0.47\linewidth]{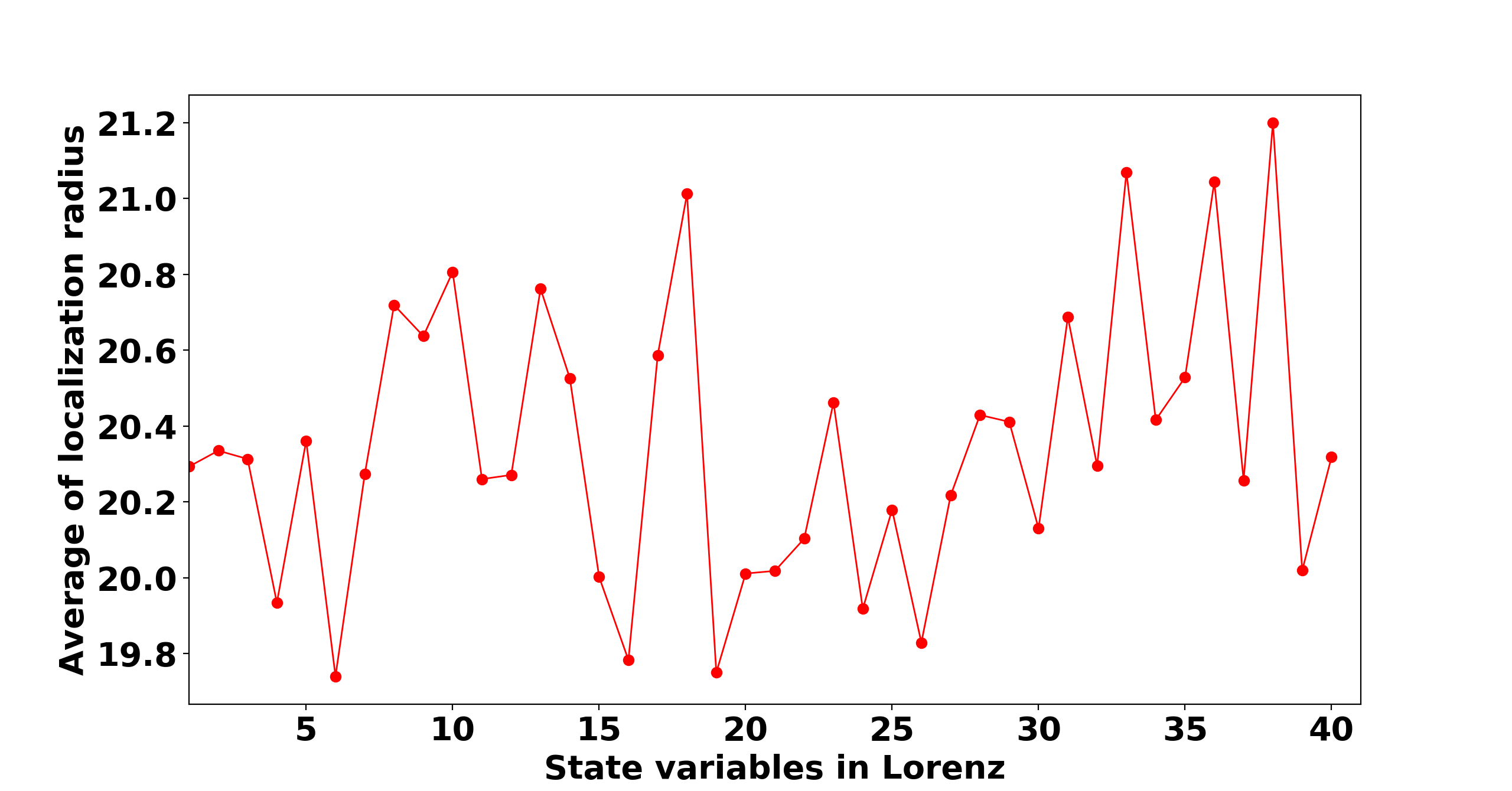}
\label{fig:lorenz_stat_change_average}
}
\subfigure[Standard deviation of localization radii]{
\includegraphics[width=0.47\linewidth]{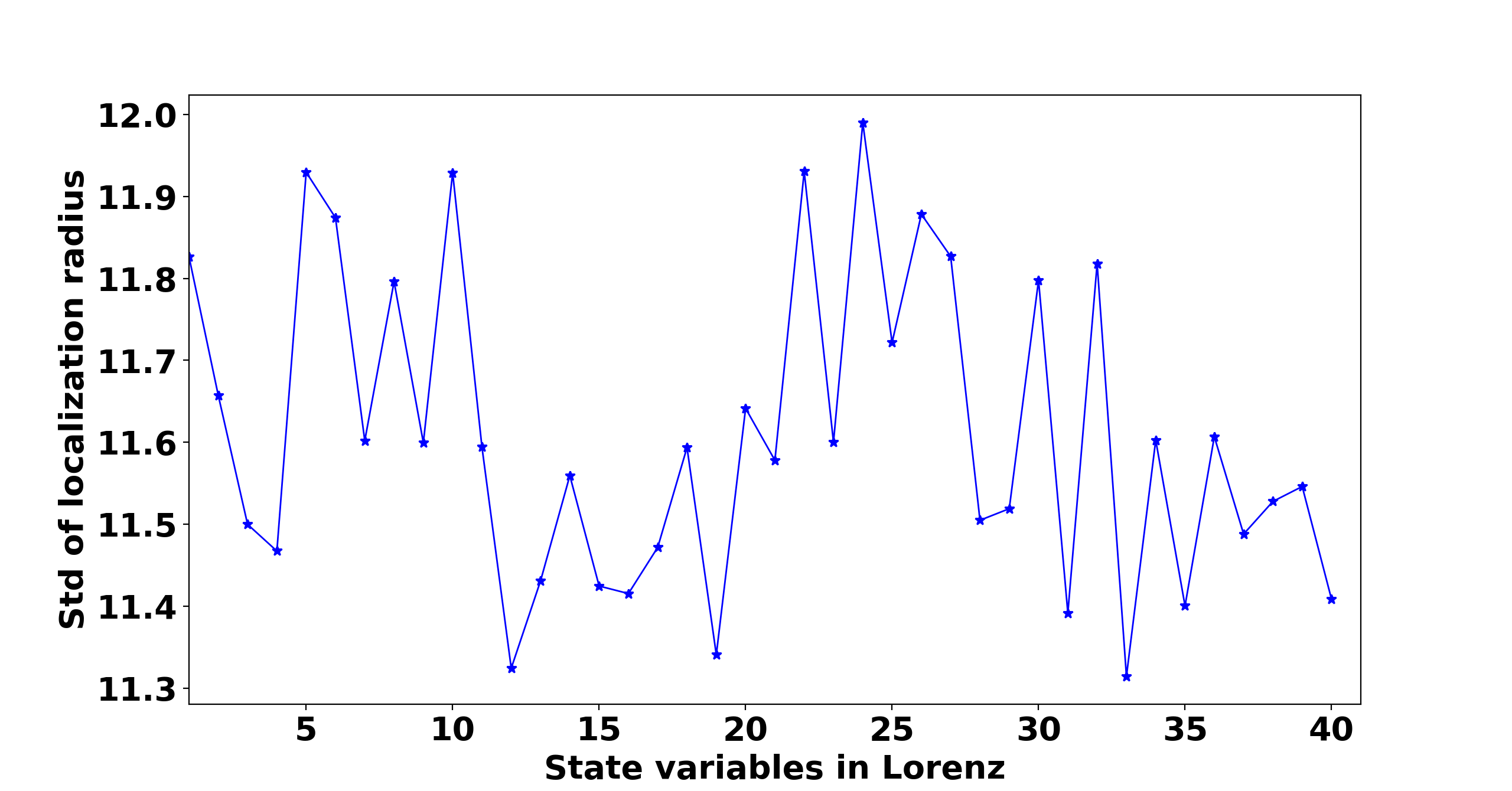}
\label{fig:lorenz_stat_change_std}
}
\caption{Statistical variability in localization radii for different state variables of the Lorenz-96 model \eqref{eq:Base_Lorenz}. The total 1000 assimilation cycles include both training and testing phases. The weights of the adaptive localization criterion are $w_1=0.7$ and  $w_2=0.3$.}
\label{fig:lorenz_stat_change}
\end{figure}

\begin{figure}[h]
\centering
\subfigure[First 100 assimilation cycles]{
\includegraphics[width=0.47\linewidth]{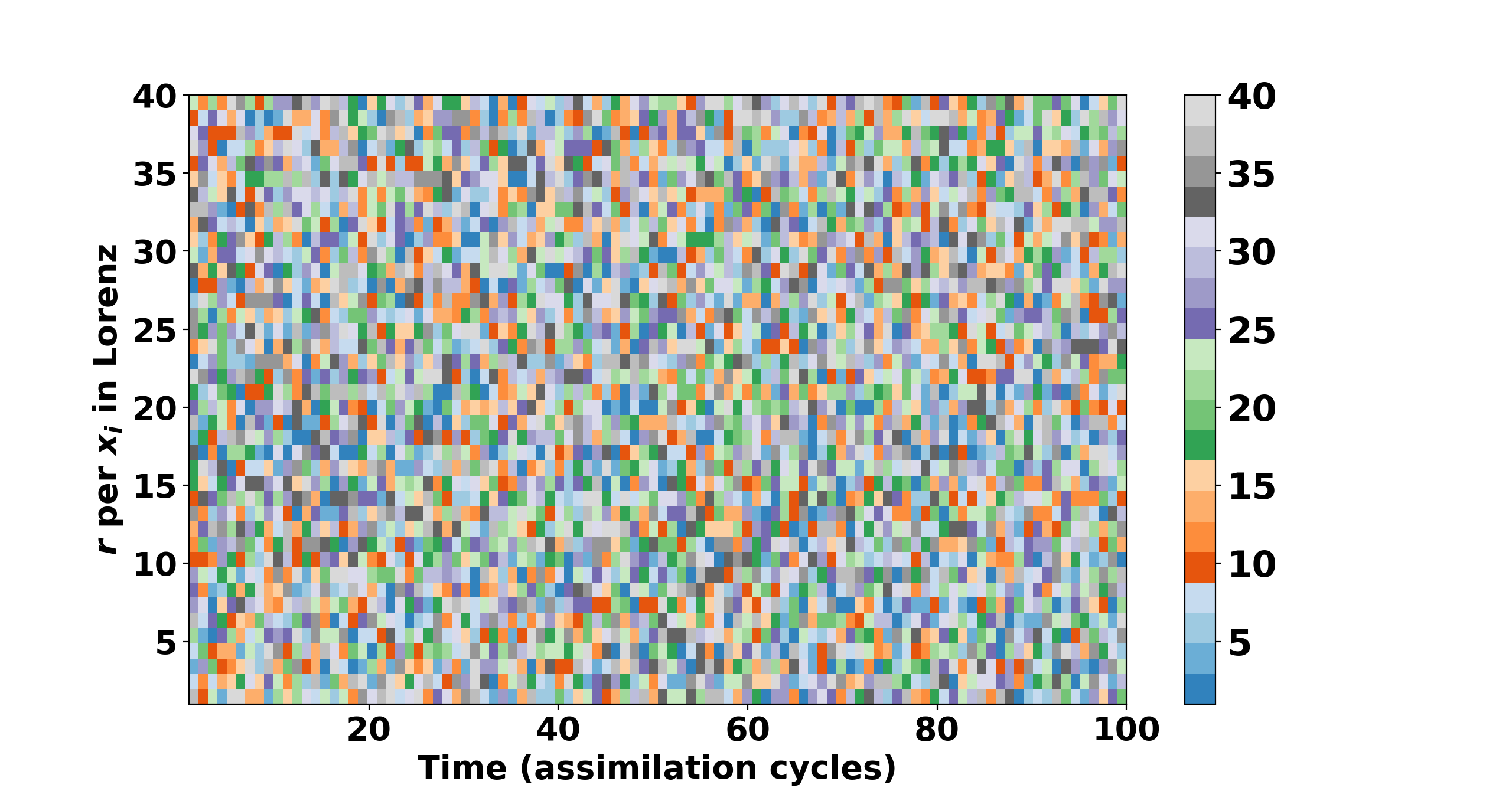}
\label{fig:lorenz_loc_change_train}
}
\subfigure[Last 100 assimilation cycles]{
\includegraphics[width=0.47\linewidth]{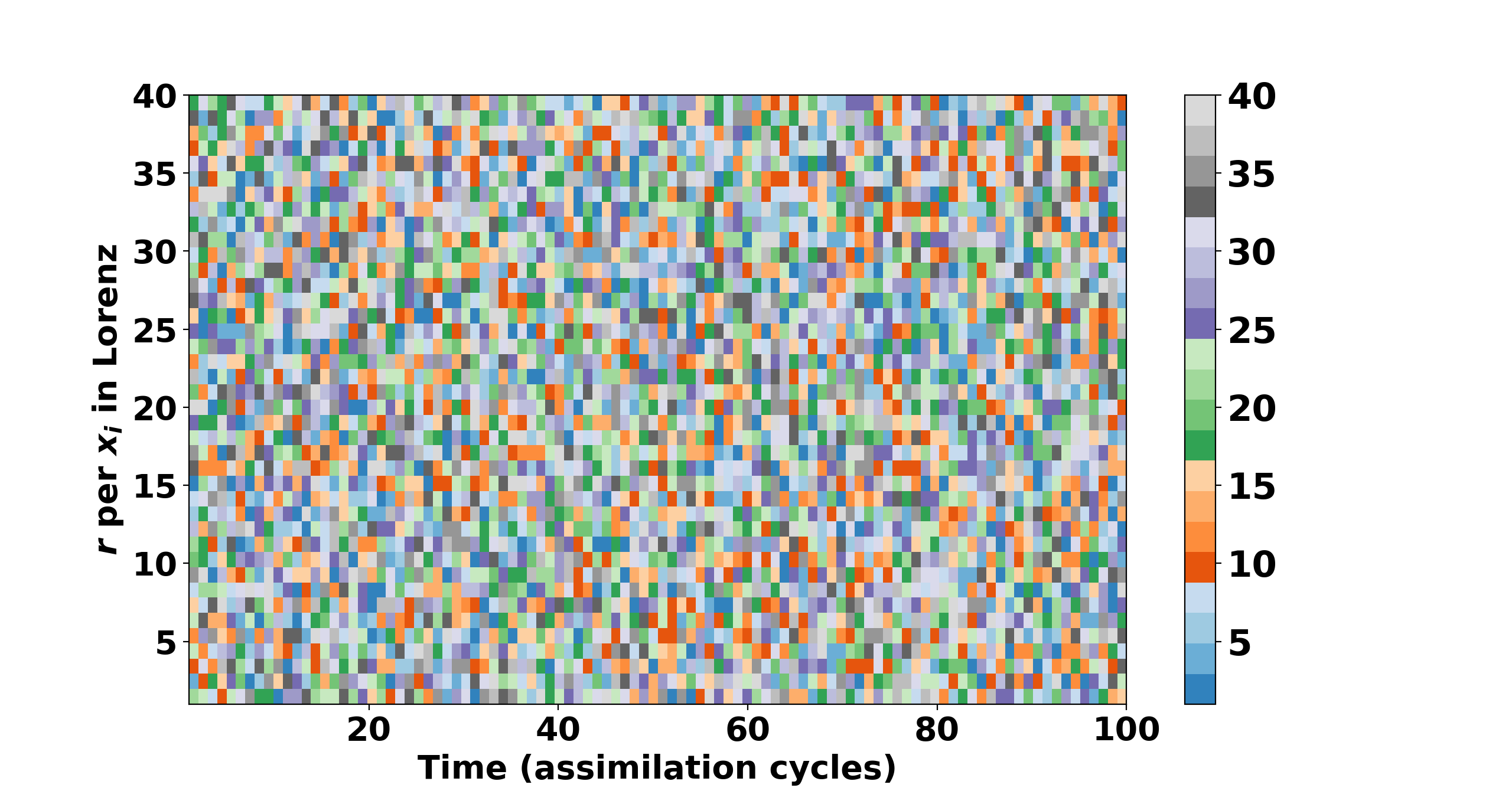}
\label{fig:lorenz_loc_change_test}
}
\caption{Evolution of localization radii in both time and space for the Lorenz-96 model \eqref{eq:Base_Lorenz}. The weights of the adaptive localization criterion are $w_1=0.7$ and $w_2=0.3$.}
\label{fig:lorenz_loc_change}
\end{figure}

Figure \ref{fig:lorenz_stat_change} shows the statistical variability in localization radii for the Lorenz model over time with the weights $w_1=0.7$ and $w_2=0.3$. Figure \ref{fig:lorenz_stat_change_average} shows the average of localization radius variability in time for each state variable of the Lorenz model and Figure \ref{fig:lorenz_stat_change_std} shows the standard deviation of localization radius change in time for each state variable. The average and standard deviations are taken over the state variables; we see that the adaptive values chosen by the algorithm can vary considerably.
This variability can be further seen in Figure \ref{fig:lorenz_loc_change}, which shows the evolution of localization radii in both time and space for the Lorenz model. The first 100 cycles of training phase and the last 100 cycles of the testing phase are selected. The weights of the adaptive localization criterion are $w_1=0.7$ and  $w_2=0.3$.
%
%

\subsection{QG model  with adaptive localization in time and space}
\label{Subsec:qg_adaptive_time_space}

The pool of localization radii for this experiment consists of random vectors of size $1085$, where each component of the vector can have values in the range of proper radii for the QG model. One practical restriction is that the localization radius used for neighboring grid points should not be too different. We noticed that having to much variability in the choice of localization radii for grid points located nearby in space may lead to physical imbalances and filter divergence. One remedy is to narrow down the range of possible radii to a limited range. Here for example, we restricted the localization radius possible values to $[2,3]$, $[3,4]$, or $[4,5]$. EnKF uses $25$ ensemble members, an inflation factor of $1.06$, and the GC localization function.

\begin{figure}[h]
\centering
\subfigure[Training phase]{
\includegraphics[width=0.47\linewidth]{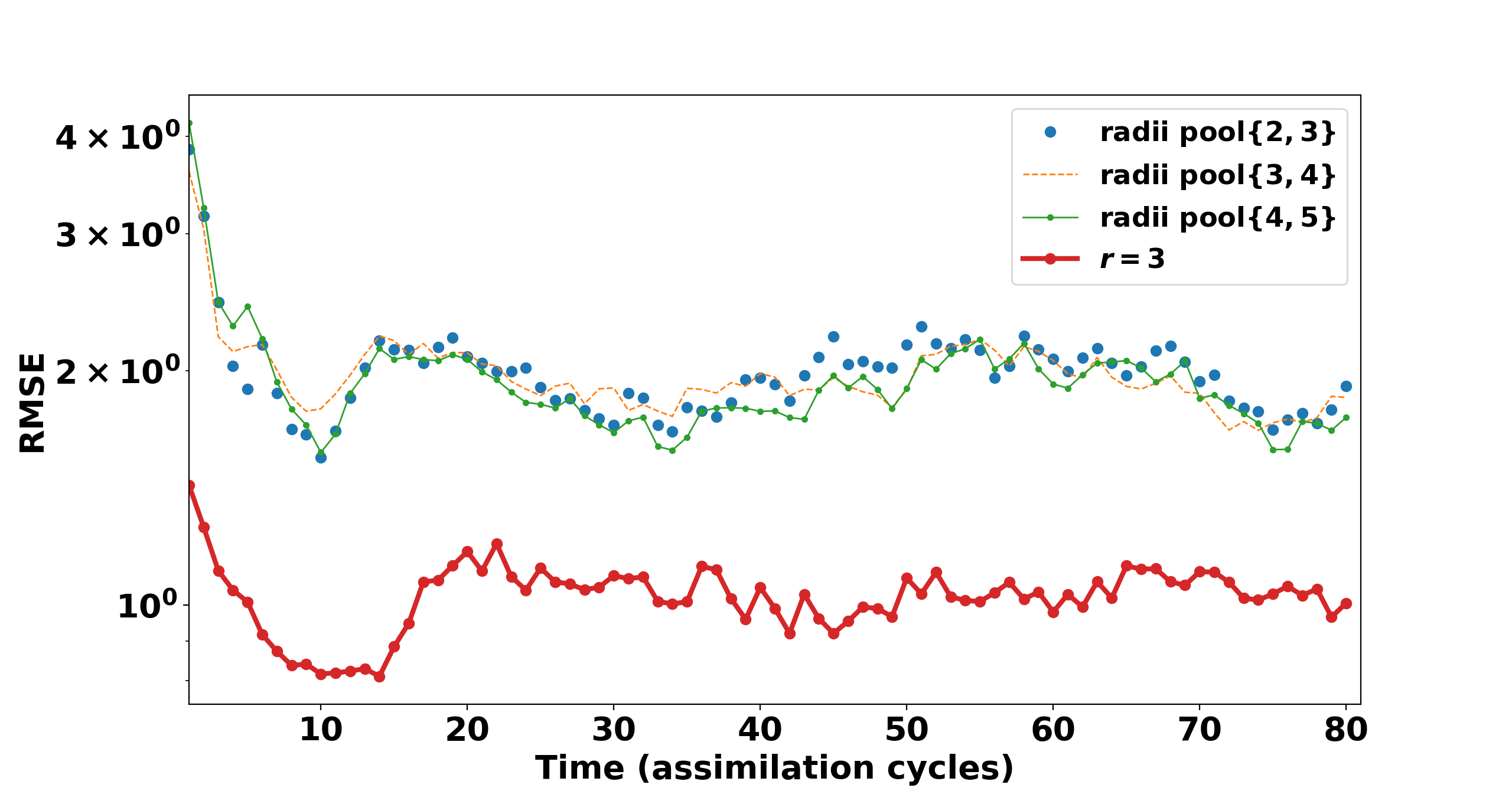}
\label{fig:QG_short_changing_train}
}
\subfigure[Testing phase]{
\includegraphics[width=0.47\linewidth]{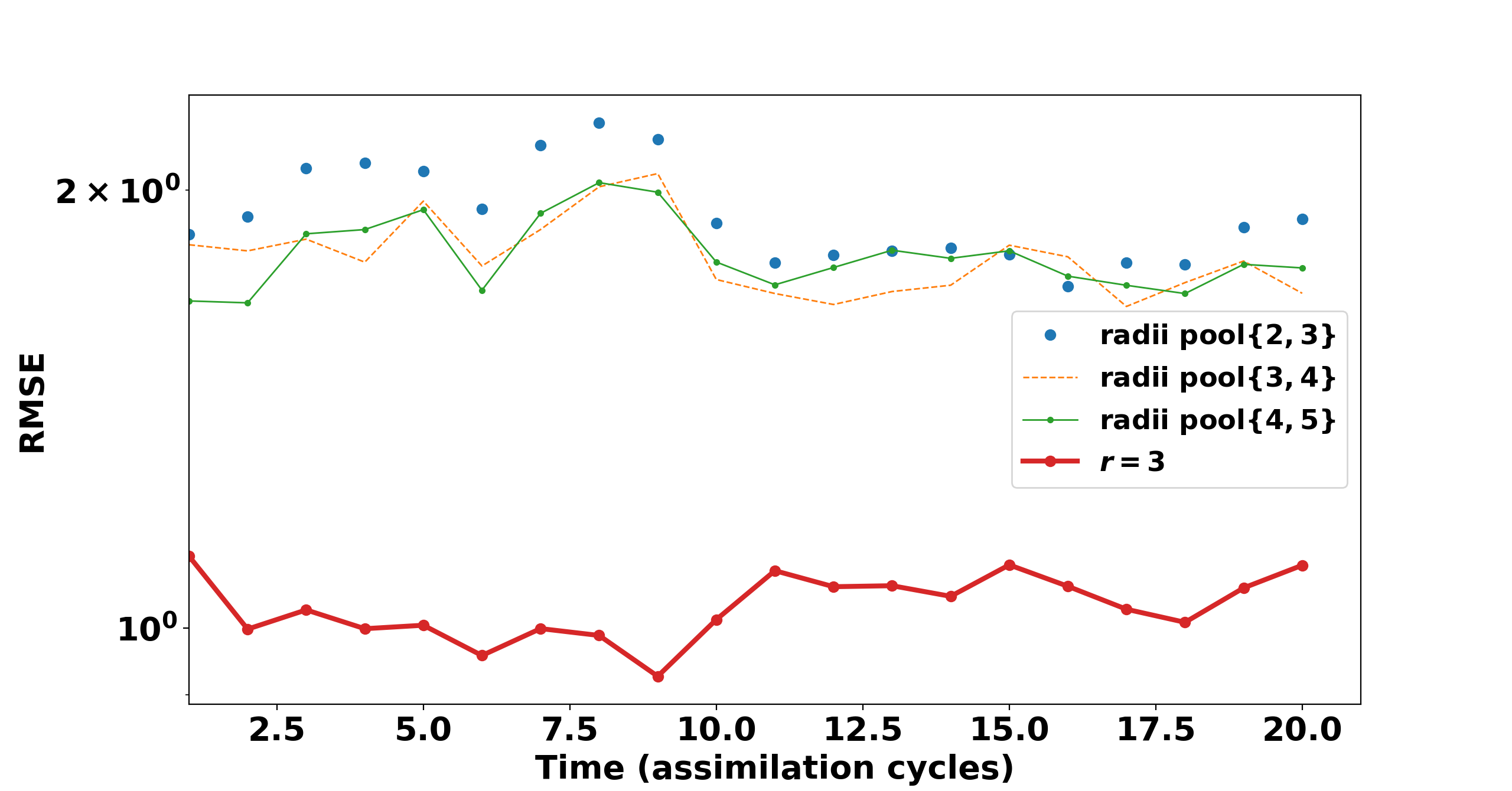}
\label{fig:QG_short_changing_test}
}
\caption{Data assimilation with the QG model \eqref{eqn:QG_Model}. EnKF is applied with a GC covariance localization function. The (\textit{log}) RMSE is shown on the vertical axis, and the time (assimilation cycles) is shown on the horizontal axis. EnKF results with adaptive covariance localization are shown for different choices of the weighting factors $w_1$, $w_2$. The localization is adaptive in time, and is compared to results with a fixed localization radius. The training phase consists of 80 assimilation cycles, and  testing phase follows with 20 assimilation cycles. The time and space adaptive radii results are not as good as those obtained with the fixed, hand-tuned radius.}
\label{fig:QG_short_changing}
\end{figure}

Figure \ref{fig:QG_short_changing} shows the RMSE of the analysis error at each assimilation cycle. The time and space adaptive radii results are not as good as those obtained with the fixed, hand-tuned radius. This is likely due to the very limited range of radii that the algorithm was allowed to test in each experiment.

\begin{figure}[h]
\centering
\subfigure[First 10 assimilation cycles]{
\includegraphics[width=0.47\linewidth]{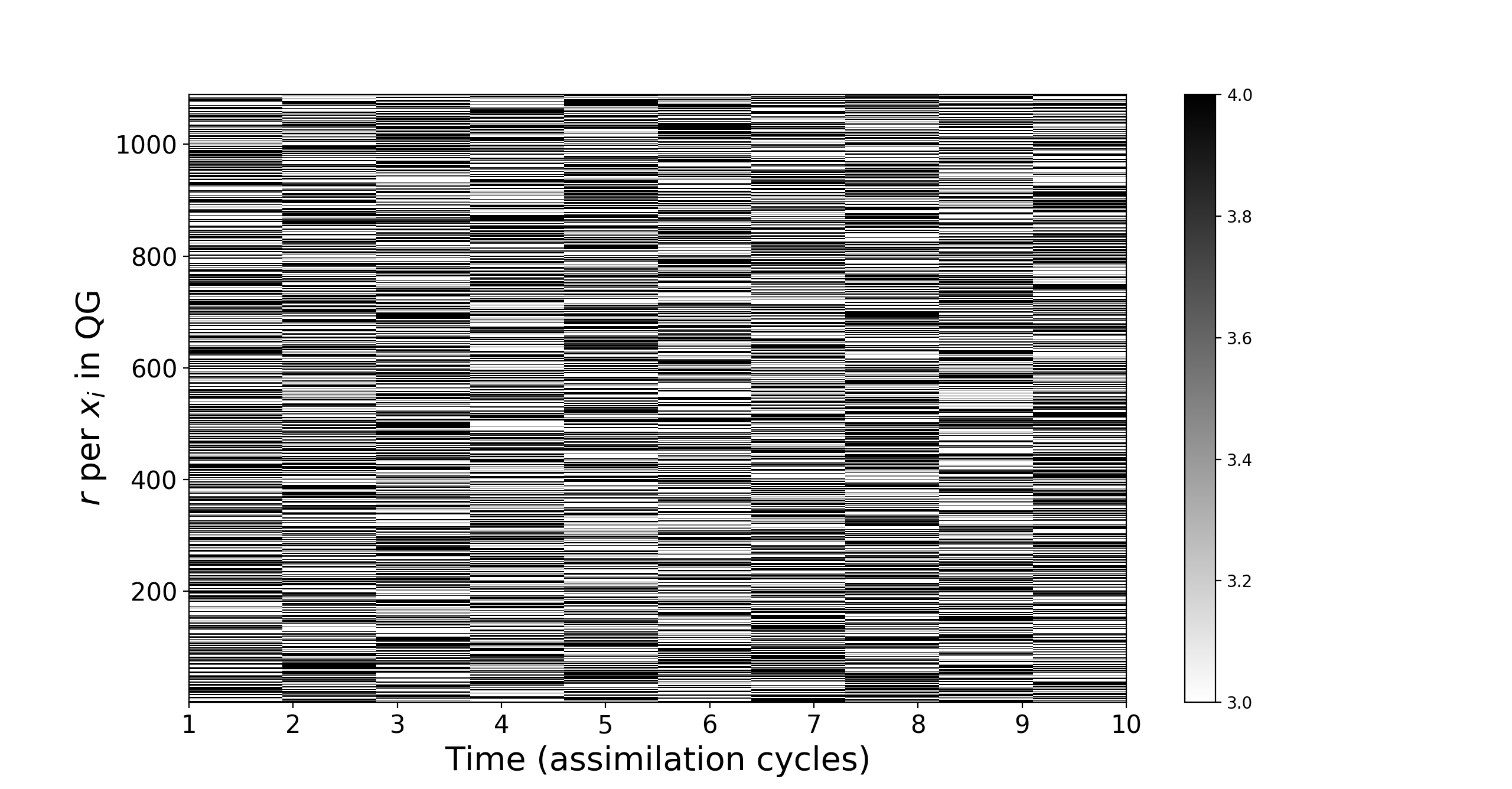}
\label{fig:QG_loc_change_train}
}
\subfigure[Last 10 assimilation cycles]{
\includegraphics[width=0.47\linewidth]{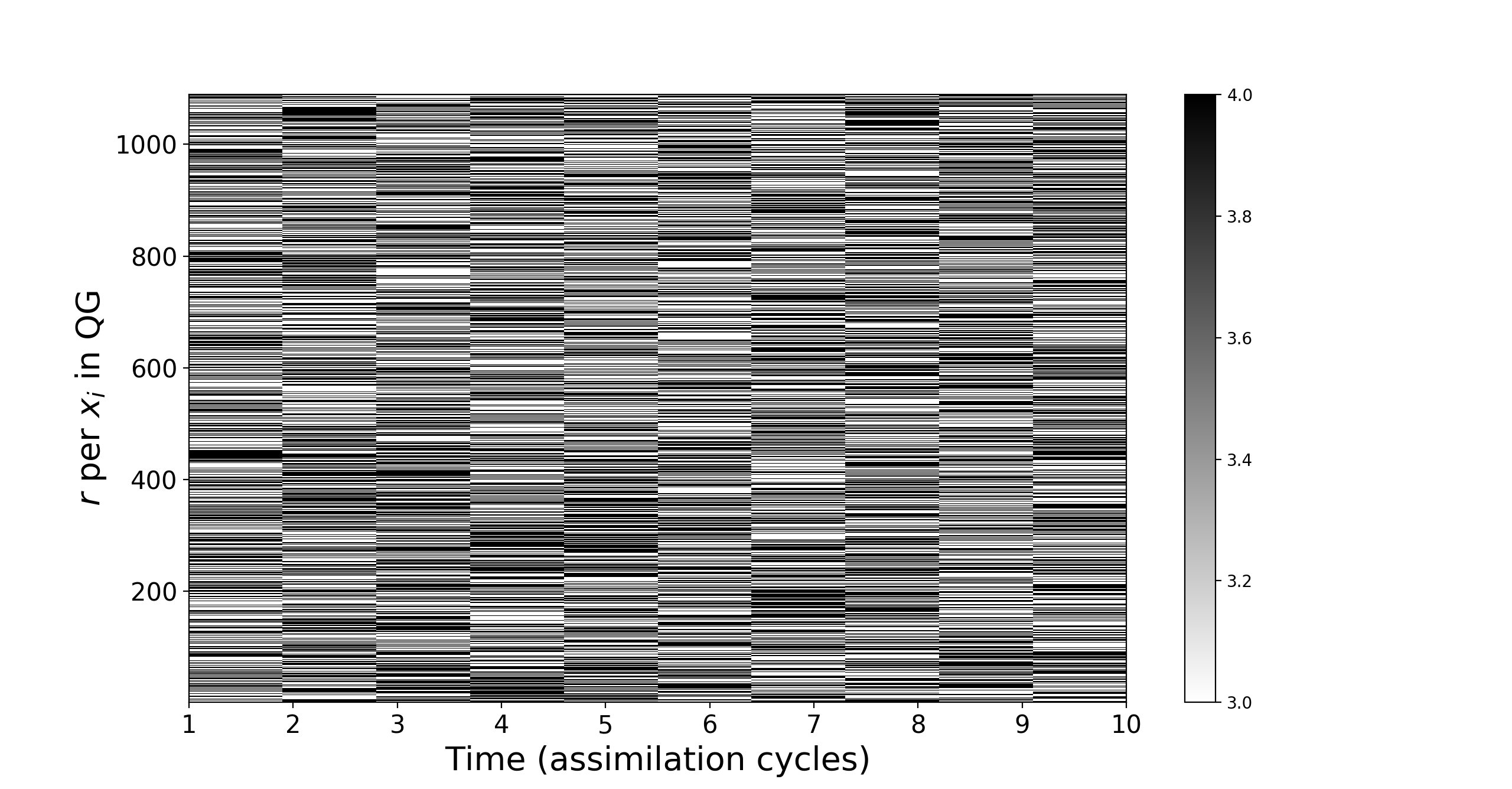}
\label{fig:QG_loc_change_test}
}
\caption{Variability of localization radius in both time and space for QG model \eqref{eqn:QG_Model}. The pool of radii is $\{3, 4\}$ The weights of the adaptive localization criterion are $w_1=0.7$ and $w_2=0.3$.}
\label{fig:QG_loc_change}
\end{figure}

Figure \ref{fig:lorenz_stat_change} shows the statistical variability in localization radii for the QG model, with the adaptive criterion weights $w_1=0.7$ and $w_2=0.3$ . The variability is computed across all state vector components. The limited range of values from which the radius selection is made leads to a small variability of the radii.
\begin{figure}[h]
\centering
\subfigure[Average of localization radius]{
\includegraphics[width=0.47\linewidth]{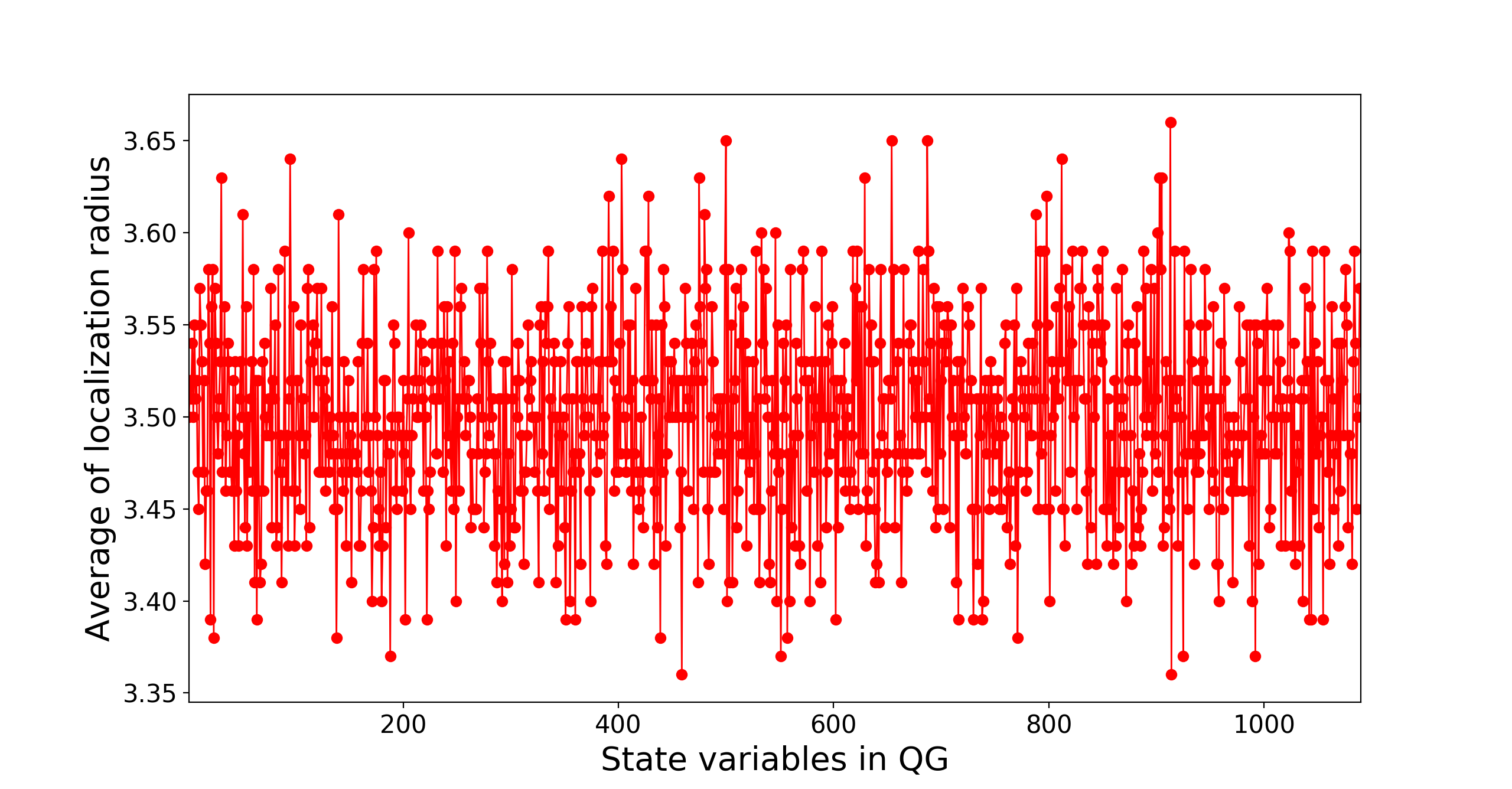}
\label{fig:QG_stat_change_average}
}
\subfigure[Standard deviation of localization radius]{
\includegraphics[width=0.47\linewidth]{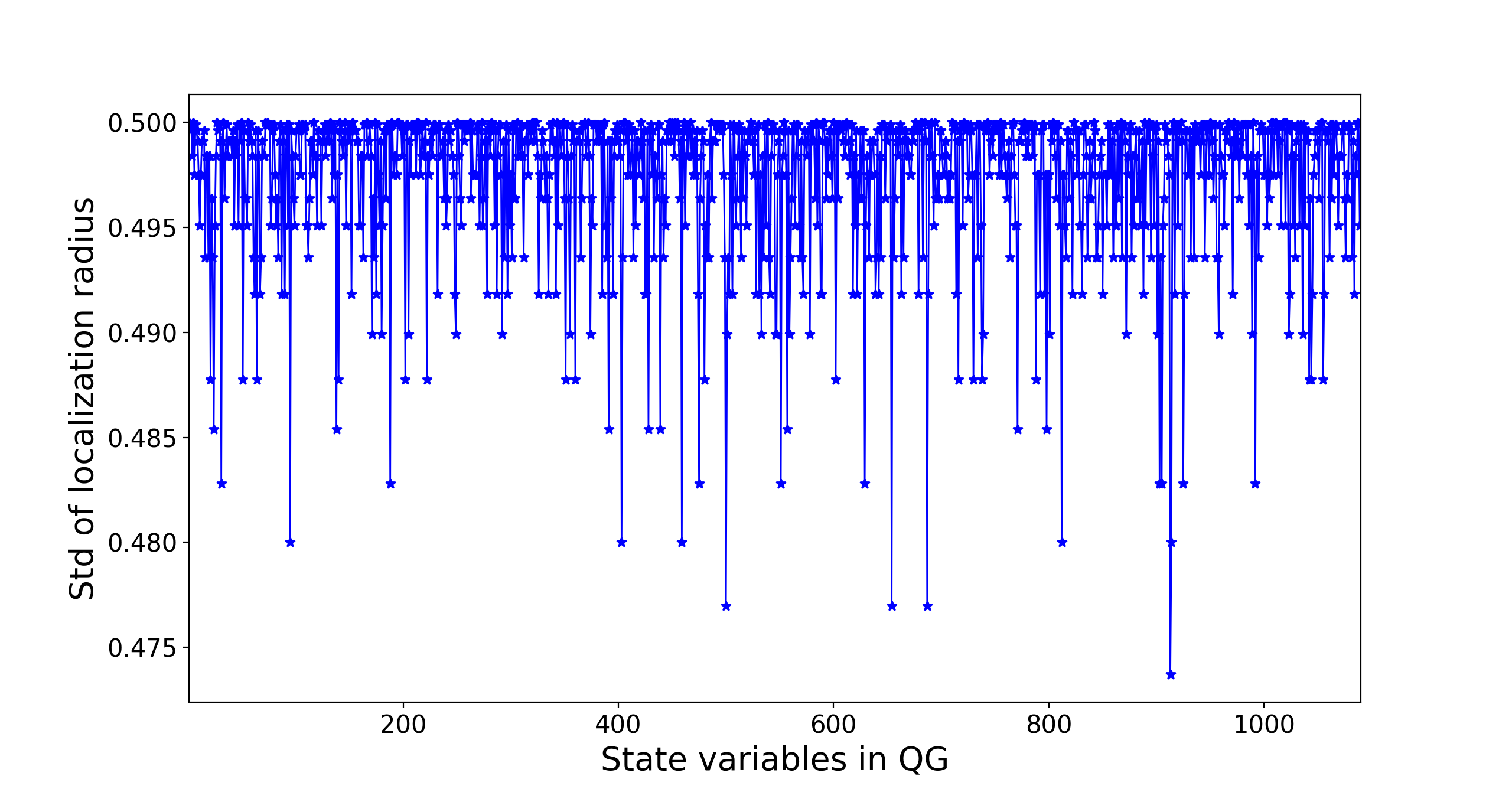}
\label{fig:QG_stat_change_std}
}
\caption{Statistical variability in the localization radius for each state variable of QG model \eqref{eqn:QG_Model}. The total assimilation cycles are 100 including both training and testing phases. The weights of the adaptive localization criterion are $w_1=0.7$ and $w_2=0.3$.}
\label{fig:QG_stat_change}
\end{figure}
The changes made by the adaptive algorithm are shown in Figure \ref{fig:QG_loc_change}  for the first 10 cycles of training phase and for the last 10 cycles of test phase. The weights of the adaptive localization criterion are $w_1=0.7$ and $w_2=0.3$. We notice that the radii chosen for different state variables seem to be uncorrelated in space or time.
%

\subsection{Discussion of numerical results}
\label{Subsec:Discussion}

Several points can be concluded from the experimental results.
Firstly, one needs to consider different decision criterion weights for different problems. Here the best weights are not the same for both models. For the Lorenz model, a combination with a larger weight for KL distance has a positive affect. A more balanced set of weights works better for the QG model.
Secondly, adaptivity leads to a considerable variability of the localization radii in both time and space. As the feature importance plots show, the values of state variables have a significant bearing on radius predictions. Moreover, the importance of all state variables is not the same, and some variables in the model  have a higher impact on the prediction of localization radii. 
Finally, the training of the localization algorithms in both time and space with the current methodology is computationally expensive. Future research will focus on making the methodology truly practical for very large models. 

\section{Concluding Remarks and Future Work}\label{Sec:Conclusions}
%
This study proposes an adaptive covariance localization approach for the EnKF family of data assimilation methods.
	Two methodologies are presented and discussed, namely adaptivity in time and adaptivity in space and time.
	The adaptive localization approach is based on random forests, a machine learning regression technique.
	The learning model can be trained off-line using historical records, e.g., reanalysis data.
    Once it is successfully trained, the regression model is used to estimate the values of localization radii in future assimilation cycles.
	Numerical results carried out using two standard models suggest that the proposed automatic approach performs at least as good as the traditional EnKF with empirically hand-tuned localization parameters.

	In order to extend the use of machine learning techniques to support data assimilation, an important question that will be addressed in future research concerns the optimal choice of features in large-scale numerical models. Specifically, one has to select sufficient aspects of the model state to carry the information needed to train a machine learning algorithm.  In the same time, the size of the features vector needs to be relatively small, even when the model state is extremely large. Next, the computational expense of the training phase is due to the fact that the analysis needs to be repeated with multiple localization radii. Future work will seek to considerably reduce the computational effort by intelligently narrowing the pool of possible radii to test, and by devising assimilation algorithms that reuse the bulk of the calculations when computing multiple analyses with multiple localization radii.

\section*{Acknowledgments}
This work was supported in part by the projects AFOSR DDDAS 15RT1037 and AFOSR Computational Mathematics FA9550-17-1-0205.

\bibliographystyle{plain}
\bibliography{Bib/data_assim_general,Bib/data_assim_HMC,Bib/data_assim_kalman,Bib/DATeS_Software,Bib/comprehensive_bibliography,Bib/localization}

\end{document}

%% file: packages_and_notations.tex
\usepackage{amssymb}
\usepackage[colorlinks,bookmarksopen,bookmarksnumbered,citecolor=red,urlcolor=red]{hyperref}
\usepackage{moreverb,dblfloatfix}
\usepackage{algorithm}
\usepackage{algpseudocode}
\usepackage{amssymb}
\usepackage{graphicx,color}
\usepackage{float}
\usepackage{amsmath,amssymb}
\usepackage{subfigure}
\usepackage{rotating}
\usepackage{multirow}
\usepackage{placeins}
\usepackage[normalem]{ulem}
\usepackage{appendix}
\usepackage{xspace}
\usepackage{listings}
\usepackage{natbib}
\newcommand{\obs}{ \mathbf{y} }

\renewcommand{\vec}[1]{\mathbf{#1}}
\newcommand{\mat}[1]{\mathbf{#1}}

\newcommand{\Nobs}{ {\rm N}_{\rm obs} }

\newcommand{\x}{ \mathbf{x} }

\newcommand{\xa}{ \mathbf{x}^{\rm a} }
\newcommand{\xtrue}{ \mathbf{x}^{\rm true} }

\newcommand{\xf}{ \mathbf{x}^{\rm f} }

\newcommand{\xbarf}{\overline{\textbf{x}}^{\rm f}}

\newcommand{\yk}{\mathbf{y}_k}

\newcommand{\norm}[1]{\left\Vert #1 \right\Vert}

\newcommand{\Nens}{{\rm N}_{\rm ens}}
\newcommand{\Nstate}{{\rm N}_{\rm state}}

\newcommand{\DKL}[1]{D_{KL}\bigl( #1 \bigr) }

\newcommand{\dates}{{DATeS}\xspace}

%% file: logo.tex
\thispagestyle{empty}
\setcounter{page}{0}

\begin{Huge}
\begin{center}
Computer Science Technical Report CSTR-{\tt1} \\
\end{center}
\end{Huge}
\vfil
\begin{huge}
\begin{center}
Azam Moosavi, Ahmed Attia, Adrian Sandu
\end{center}
\end{huge}

\vfil
\begin{huge}
\begin{it}
\begin{center}
``{\tt A Machine Learning Approach to Adaptive Covariance Localization}''
\end{center}
\end{it}
\end{huge}
\vfil

\begin{large}
\begin{center}
Computational Science Laboratory \\
Computer Science Department \\
Virginia Polytechnic Institute and State University \\
Blacksburg, VA 24060 \\
Phone: (540)-231-2193 \\
Fax: (540)-231-6075 \\ 
Email: \url{azmosavi@cs.vt.edu} \\
Web: \url{http://csl.cs.vt.edu}
\end{center}
\end{large}

\vspace*{1cm}

\begin{tabular}{ccc}
\includegraphics[width=2.5in]{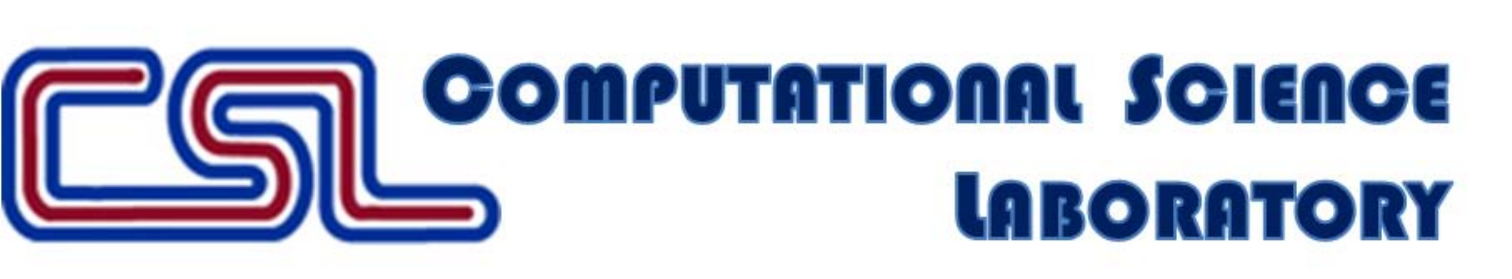}
&\hspace{2.5in}&
\includegraphics[width=2.5in]{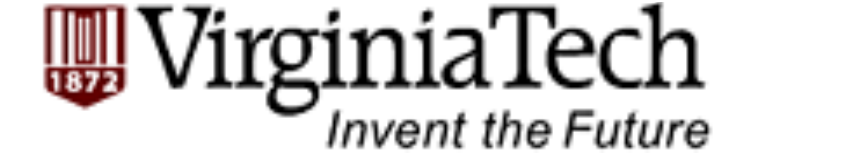} \\
{\bf\em Innovative Computational Solutions} &&\\
\end{tabular}

\newpage